\documentclass{aa}

\usepackage[english]{babel}
\usepackage{graphicx}
\usepackage[varg]{txfonts}
\usepackage{natbib}
\usepackage{longtable}
\usepackage{array}
\usepackage{multirow}
\usepackage{amsmath}

\bibpunct{(}{)}{;}{a}{}{,}
\begin{document}

\title{Millimeter wave spectrum and search for vinyl isocyanate toward Sgr B2(N) with ALMA
\thanks{Tables \ref{transitions-trans}, \ref{transitions-cis}, \ref{predictions-trans}, and \ref{predictions-cis} are only available in electronic form
at the CDS via anonymous ftp to cdsarc.u-strasbg.fr (130.79.128.5) or via http://cdsweb.u-strasbg.fr/cgi-bin/qcat?J/A+A/
}}

   \author{K.~V\'avra\inst{1}
        \and L.~Kolesnikov\'a\inst{1}
        \and A.~Belloche\inst{2}
        \and R.~T. Garrod\inst{3}
        \and J.~Kouck\'y\inst{1}
        \and T.~Uhl\'ikov\'a\inst{1}
        \and K.~Lukov\'{a}\inst{1}
        \and J.-C.~Guillemin\inst{4}
        \and P.~Kania\inst{1}
        \and H.~S.~P.~M\"uller\inst{5}
        \and K.~M.~Menten\inst{2}
        \and \v{S}. Urban\inst{1}
          }

\institute{Department of Analytical Chemistry, University of Chemistry and Technology,
Technick\'{a} 5, 166 28 Prague 6, Czech Republic\\\email{karel.vavra@vscht.cz, lucie.kolesnikova@vscht.cz}
\and
Max-Planck-Institut f\"{u}r Radioastronomie, Auf dem H\"{u}gel 69, 53121 Bonn, Germany
\and
Departments of Chemistry and Astronomy, University of Virginia, Charlottesville, VA 22904, USA
\and
Univ Rennes, Ecole Nationale Sup\'erieure de Chimie de Rennes, CNRS, ISCR UMR6226, F-35000 Rennes, France
\and
Astrophysik/I. Physikalisches Institut, Universit{\"a}t zu K{\"o}ln, Z{\"u}lpicher Str. 77, 50937 Cologne, Germany}

\date{Received ; accepted }

\titlerunning{The rotational spectrum and ISM search of vinyl isocyanate}
\authorrunning{V\'avra et al.}


\abstract
   {The interstellar detections of isocyanic acid (HNCO), methyl isocyanate (CH$_3$NCO), and very recently also ethyl isocyanate (C$_2$H$_5$NCO), open the question of the possible detection of vinyl isocyanate (C$_2$H$_3$NCO) in the interstellar medium. There are only low-frequency spectroscopic data (<~40~GHz) available for this species in the literature. This makes predictions at higher frequencies rather uncertain hampering its search in space by millimeter wave astronomy.}
   {The aim of the present study is, on one hand, to extend the laboratory rotational spectrum of vinyl isocyanate into the millimeter wave region and, on the other hand, to undertake a first check for its presence in the high-mass star forming region Sgr B2, where other isocyanates and a plethora of complex organic molecules are observed.
      }
   {The pure rotational spectrum of vinyl isocyanate was recorded in the frequency regions 127.5--218 and 285--330~GHz using the Prague millimeter wave spectrometer. The spectral analysis was supported by high-level quantum-chemical calculations. On the astronomy side, we assumed local thermodynamic equilibrium to compute synthetic spectra of
vinyl isocyanate and to search for it in the ReMoCA
survey performed with the Atacama Large Millimeter/submillimeter Array
(ALMA) toward the high-mass star forming protocluster Sgr~B2(N). Additionally, we searched for the related molecule ethyl isocyanate in the same source.}
   {Accurate values for the rotational and centrifugal distortion constants are reported for the ground vibrational states of \textit{trans} and \textit{cis} vinyl isocyanate from the analysis of more than 1000 transitions. We report nondetections of vinyl and ethyl isocyanate toward the main hot
core of Sgr B2(N). We find that vinyl and ethyl isocyanate are at least 11
and 3 times less abundant than methyl isocyanate in this source,
respectively.
}
   {Although the precise formation mechanism of interstellar methyl
isocyanate itself remains uncertain, we infer from existing astrochemical
models that our observational upper limit for the CH$_3$NCO:C$_2$H$_5$NCO
ratio in Sgr B2(N) is consistent with ethyl isocyanate being formed on
dust grains via the abstraction or photodissociation of an H atom from
methyl isocyanate, followed by the addition of a methyl radical. The
dominance of such a process for ethyl isocyanate production, combined with
the absence of an analogous mechanism for vinyl isocyanate, would indicate
that the ratio C$_2$H$_3$NCO:C$_2$H$_5$NCO should be rather less than
unity. Even though vinyl isocyanate was not detected toward Sgr B2(N), results of this work represent a
significant improvement on previous low-frequency studies and will allow the astronomical community to continue in
searching for this species in the universe.}


   \maketitle
%

\section{Introduction}
\label{sect_intro}

      \begin{figure*}[ht]
   \centering
   \includegraphics[trim = 45mm 5mm 35mm 10mm, clip, width=16.5cm]{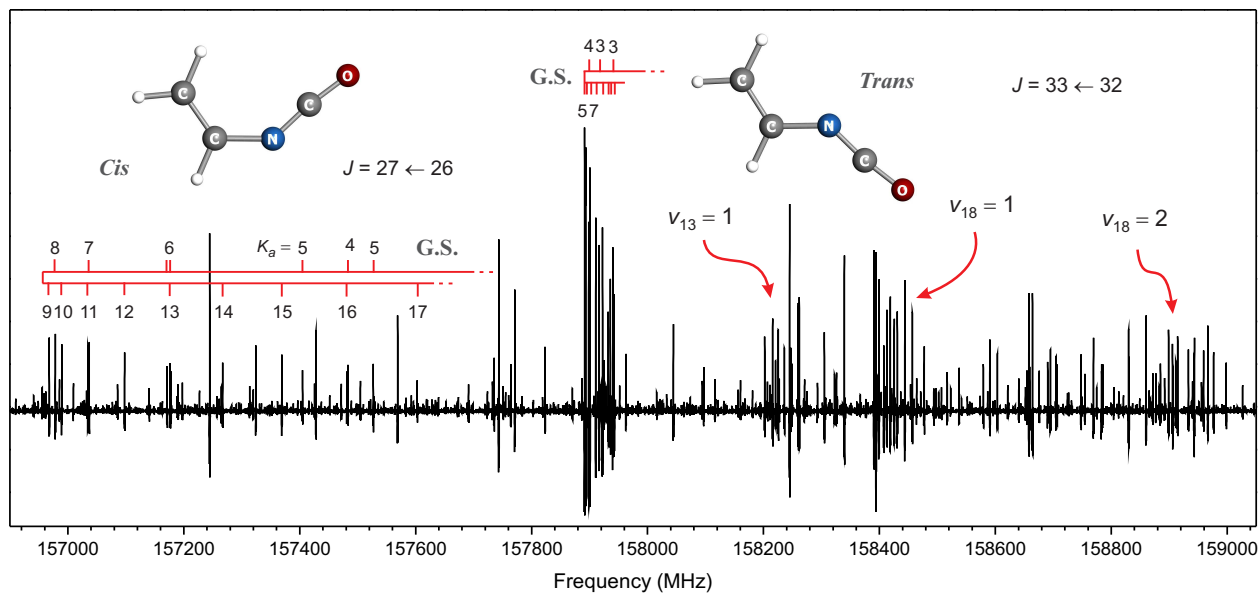}
      \caption{Two isomers of vinyl isocyanate and their characteristic features in the room-temperature millimeter wave spectrum. The spectrum reveals compact groups of $a$-type $R$-branch transitions of the \textit{trans} species in the ground state and $v_{18}=1$, $v_{13}=1$, and $v_{18}=2$ excited vibrational states. Weaker $a$-type $R$-branch transitions of the less stable \textit{cis} isomer are at the present scale observed in the form of significantly more diffuse groups.}
      \label{spectrum}
   \end{figure*}

A few isocyanate bearing molecules have been detected in the interstellar medium (ISM).
The simplest one, isocyanate radical (NCO), has been observed relatively recently in the line survey of L483 \citep{Marcelino2018},
while isocyanic acid (HNCO) belongs among the early molecules detected in the universe dating back to 1972 \citep{Snyder1972,Buhl1972}. HNCO is a well established interstellar molecule observed in the direction of a variety of sources such as TMC-1 \citep{Brown1981}, IRAS~16293-2422 \citep{Bisschop2008,Coutens2016}, L483 \citep{Marcelino2018}, NGC~6334I \citep{Ligterink2020}, G31.41+0.31 \citep{Colzi2021}, Serpens~SMM1 \citep{Ligterink2021}, and G331.512-0.103 \citep{Canelo2021}.
Its protonated form, H$_2$NCO$^+$, had been tentatively observed in Sgr~B2 by \cite{Gupta2013}, but later on its presence in space has been confirmed by \cite{Marcelino2018}.
H$_2$NCO$^+$ has been recently detected toward the molecular cloud G+0.693-0.027 \citep{Rodriguez-Almeida21}
where NCO, cyanogen isocyanate (NCNCO), and ethynyl isocyanate (HCCNCO) were also searched for.
Methyl derivative of HNCO, methyl isocyanate (CH$_3$NCO), has been first detected in the Sgr~B2(N) and Orion~KL star-forming regions \citep{Halfen15,Cernicharo16,Belloche17}. Further observations have shown that it is also present in other sources such as IRAS~16293-2422 \citep{Ligterink17,Martin-Domenech2017}, G31.41+0.31 \citep{Gorai2021,Colzi2021}, Serpens~SMM1 \citep{Ligterink2021}, G10.47+0.03 \citep{Gorai2020}, G+0.693-0.027 \citep{Zeng2018}, and G328.2551-0.5321 \citep{Csengeri2019}. Very recently, the even more complex ethyl isocyanate (C$_2$H$_5$NCO) has been discovered toward the G+0.693-0.027 molecular cloud by \cite{Rodriguez-Almeida21}. Isocyanates are therefore quite widespread across the Galaxy and more such species might be detected in future. In addition, new molecular discoveries are reinforced by increasing the sensitivity and detector bandwidths of astronomical observing capabilities \citep{Jorgensen2020,Tercero2021} as demonstrated by some of the latest astronomical detections \citep{Cernicharo2021,Rivilla2021,McCarthy2021} which contributed to the recent huge jump in the total number of detected compounds \citep{McGuire2022}.

The aforementioned discoveries opened unique possibilities to uncover the fundamentals of the chemistry of isocyanates in space and their possible role as precursors in the formation of other compounds \citep{Majumdar2017,Zeng2018}. For example, H$_2$NCO$^+$ has been considered as a candidate reactant partner of HC$_3$N in the synthesis of nucleobases \citep{Choe2021} and CH$_3$NCO as a precursor for the synthesis of  N-methylformamide (CH$_3$NHCHO, \citealt{Belloche17}). The latter has been observed toward Sgr~B2 and NGC~6334I star-forming regions \citep{Belloche17,Belloche19,Ligterink2020} and very recently also toward the hot core G31.41+0.31 \citep{Colzi2021}. This possible link with prebiotic chemistry further increases an interest in isocyanates as targets for laboratory spectroscopic studies in view of future observations.

In the present work, the molecule in question is vinyl isocyanate (C$_2$H$_3$NCO),
the most thermodynamically stable isomer with a C$_3$H$_3$NO formula \citep{Fourre2020}.
It is more complex than CH$_3$NCO and at the same time it contains two hydrogen atoms fewer than recently detected C$_2$H$_5$NCO. In addition, it bears in its backbone the vinyl functional group which is a common structural motif in several other interstellar compounds such as vinyl cyanide \citep[C$_2$H$_3$CN;][]{Gardner1975}, vinylacetylene \citep[C$_2$H$_3$CCH;][]{Cernicharo2021c}, vinylcyanoacetylene \citep[C$_2$H$_3$C$_{3}$N;][]{Kelvin_Lee_2021}, and vinylamine \citep[C$_2$H$_3$NH$_{2}$;][]{Zeng2021}. A search for vinyl isocyanate in the line survey of the G+0.693-0.027 molecular cloud has been reported very recently \citep{Rodriguez-Almeida21} on the basis of the only rotational spectrum reported so far below 40~GHz \citep{Bouchy1977,Kirby1978,Bouchy1979}. On the other hand, no such trials have been performed in the millimeter wave surveys of other interstellar sources very likely due to the lack of laboratory information in the millimeter wave spectral region.

The aim of the present work is therefore two-fold. First, we want to extend the laboratory rotational spectrum of vinyl isocyanate up to 330 GHz and analyze the ground state rotational transitions of its two stable planar forms: \textit{trans} and \textit{cis} (see Fig. \ref{spectrum}). The latter has been found by \cite{Kirby1978} to be less stable by 4.99(24)~kJ~mol$^{-1}$ or 417(20)~cm$^{-1}$ or 600(29)~K, where the numbers in parentheses represent uncertainties in units of the last decimal digits.
These new measurements and analyses then lay a foundation to accomplish a search for interstellar signatures of vinyl isocyanate by millimeter wave astronomy.
We target the high-mass star-forming region Sgr B2(N) which reveals an astonishingly rich collection of molecular species, including isocyanates.

\begin{table*}
\caption{Spectroscopic constants for the \textit{trans} and \textit{cis} isomers of vinyl isocyanate in their ground vibrational states ($A$-reduction, I$^{\text{r}}$-representation) in comparison with previously published results and quantum-chemical calculations.}
\label{constants}
\begin{center}
\begin{footnotesize}
\setlength{\tabcolsep}{6.0pt}
\begin{tabular}{ l r r r r r r r}
\hline\hline
\vspace{-0.3cm}\\
 & \multicolumn{3}{c}{\textit{Trans}} &  & \multicolumn{3}{c}{\textit{Cis}}  \\
\cline{2-4}
\cline{6-8}
\vspace{-0.2cm}\\
 &  This work  & \cite{Kirby1978} &  Calculated\tablefootmark{a} &   &  This work  & \cite{Kirby1978} &  Calculated\tablefootmark{a} \\
\hline
\vspace{-0.2cm}\\
$A            $  /MHz             & 62586.3098 (25)\tablefootmark{b}  & 62584.051 (35) &   62314.45     &  &  20144.090 (41)         &  20146.8 (10)    &  20193.42     \\
$B            $  /MHz                       &      2437.747011 (86)   &   2437.730 (3) &    2438.70     &  &   3107.45256 (44)       &   3107.267 (20)  &  3096.35      \\
$C            $  /MHz                       &      2346.477545 (88)   &   2346.507 (1) &    2346.96     &  &   2689.42677 (34)       &   2689.513 (25)  &  2682.28      \\
$\mathit{\Delta_{J}}   $  /kHz                       &         0.266716 (35)   &    0.321 (9)   &   0.2695       &  &      3.03750 (13)       &  2.23 (26)       &   3.078       \\
$\mathit{\Delta_{JK}}  $  /kHz                       &      --14.3999 (20)     &    --14.30 (7) &  --18.48       &  &   --80.8627 (28)        & --80.26 (71)     & --80.49       \\
$\mathit{\Delta_{K}}   $  /kHz                       &      2270.62 (24)       &      ...       &   2343         &  &    841.7 (15)           &     ...          &   780.0       \\
$\delta_{J}   $  /kHz                       &         0.0182857 (54)  &      ...       &  0.02243       &  &      0.830560 (99)      &     ...          &  0.8380       \\
$\delta_{K}   $  /kHz                       &        10.3755 (90)     &      ...       &  8.657         &  &     13.792 (22)         &     ...          &  11.62        \\
$\mathit{\Phi_{J}}     $  /Hz                        &      0.0002564 (41)     &      ...       &  0.0002777     &  &      0.008427 (26)      &     ...          &  0.009854     \\
$\mathit{\Phi_{JK}}    $  /Hz                        &    --0.04791 (43)       &      ...       & --0.07105      &  &      0.2450 (49)        &     ...          &  0.2170       \\
$\mathit{\Phi_{KJ}}    $  /Hz                        &    --2.129 (69)         &      ...       & --2.487        &  &   --14.050 (32)         &     ...          & --12.44       \\
$\mathit{\Phi_{K}}     $  /Hz                     & --471.8\tablefootmark{c}   &      ...       &   --471.8      &  &     289 (24)            &     ...          &  130.8        \\
$\phi_{J}     $  /mHz                        &      0.03926 (87)    &      ...       &  0.06060    &  &      3.583 (16)      &     ...          &  4.221     \\
$\phi_{JK}    $  /Hz                        &      0.0311 (20)        &      ...       &   0.01395      &  &    --0.0271 (43)        &     ...          &  0.01708      \\
$\phi_{K}     $  /Hz                    & 21.58\tablefootmark{c}      &      ...       &  21.58         &  &     12.40 (36)          &     ...          &   10.60       \\
$  L_{JK}     $  /mHz                       &         ...             &      ...       &     ...        &  &   --0.16472 (85)        &     ...          &   ...         \\
$ L_{KKJ}     $  /mHz                       &         ...             &      ...       &     ...        &  &     5.267 (96)          &     ...          &   ...         \\
$ P_{KKKJ}    $  /mHz                       &         ...             &      ...       &     ...        &  &   --0.00169 (11)        &     ...          &   ...         \\

$\Delta E_{\text{ZPE}} $\tablefootmark{d}  /cm$^{-1}$  &  ...         &      ...       &     0          &  &         ...             &     417 (20)     &   301          \\
$J_{\text{min}}/J_{\text{max}}$             &     4 / 80              &      4 / 19    &     ...        &  &  1 / 61  &  4 / 6  & ... \\
$K_{a}^{\text{min}}/K_{a}^{\text{max}}$     &     0 / 7               &      0 / 7     &     ...        &  &  0 / 21  &  0 / 5  & ... \\
$N$\tablefootmark{e}                        &     464                 &       30       &     ...        &  &  608     &   18    & ... \\
$\sigma_{\text{fit}}$\tablefootmark{f}/MHz  &     0.024               &      ...       &     ...        &  &  0.030   &   ...   & ... \\
$\sigma_{\text{w}}$\tablefootmark{g}        &     0.90                &      ...       &     ...        &  &  0.88    &   ...   & ... \\
\hline
\end{tabular}
\end{footnotesize}
\end{center}
\tablefoot{
\tablefoottext{a}{Calculated at CCSD/cc-pVTZ level of theory.}
\tablefoottext{b}{The numbers in parentheses are the parameter uncertainties in units of the last decimal digits. Their values are close to 1$\sigma$ standard uncertainties (67\% confidence level) because the unitless (weighted) deviation of the fit is close to 1.0. SPFIT/SPCAT program package \citep{Pickett1991} was used for the analysis.}
\tablefoottext{c}{Fixed to the calculated value, which is usually a preferred constraint over the zero or poorly determined value \citep{Urban1990,Koucky2013}.}
\tablefoottext{d}{Relative energy with respect to the global minimum, taking into account the zero-point energy (ZPE).}
\tablefoottext{e}{Number of distinct frequency lines in the fit.}
\tablefoottext{f}{Root mean square deviation of the fit.}
\tablefoottext{g}{Unitless (weighted) deviation of the fit.}}
\end{table*}


\section{Experimental details}
{\label{s:experiments}}

\subsection{Synthesis}

The sample of vinyl isocyanate was prepared by Curtius rearrangement of acryloyl azide under vacuum (0.1~mbar) using a modified synthesis of \cite{Kirby1978}.
Briefly, sodium azide NaN$_{3}$ (3.25~g, 50~mmol) was mixed with 30~mL of diethylene glycol dibutyl ether in a three-necked flask equipped
with a magnetic stirring bar and a stopcock. The flask was evacuated to a pressure of about 0.1~mbar. Keeping the stopcock closed, the flask was immersed in a bath at --20~$^{\circ}$C. In the next step, acryloyl chloride (2.85~g, 31~mmol) was added in five portions (5 x 500~$\mu$L) through a septum. The mixture was stirred for two hours at 0~$^{\circ}$C allowing
the formation of acryloyl azide. The stopcock was then opened and all volatiles passed into the vacuum line containing a quartz tube in an oven heated approximately to 500~$^{\circ}$C and two successive cold U-tubes. The first U-tube was immersed in a bath at --70~$^{\circ}$C to remove impurities and the second one in liquid nitrogen bath to collect
vinyl isocyanate. The final product was used without any further purification. The main advantage of this experimental procedure was to avoid isolation of the potentially explosive acryloyl azide.

\subsection{Spectroscopic measurements}

The rotational spectrum of vinyl isocyanate was recorded in the frequency regions 127.5--218 and 282--330~GHz using the upgraded Prague semiconductor millimeter wave spectrometer. The spectrometer is based on a sequential multiplication of the fundamental synthesizer frequency (lower than 50 GHz) by a set of active and passive multipliers and a phase-sensitive detection as described in \cite{Kania2006}. The 2.8 and 2.3 meters long Pyrex glass free-space cells were used for the measurements. The optical path lengths were doubled to 5.6 and 4.6~m by roof-top mirrors. The millimeter-wave radiation was modulated at the modulation frequency of 28 kHz
and the detected signal was demodulated by means of a lock-in amplifier working at twice the modulation frequency.
All spectra were registered by upward and downward frequency scanning and averaged. The sample was kept at room temperature and a pressure of around 20~$\mu$bar during the experiments.

\section{Quantum-chemical calculations}
{\label{s:theory}}

Although vinyl isocyanate has been the subject of some computational studies \citep{Badawi2001,Olsen1979}, we have undertaken our own calculations
in order to obtain a reasonable estimation for the spectroscopic parameters relevant to this work.
We used coupled-cluster approximation on the level of coupled-cluster-single-double (CCSD) model \citep{ccsd}
as it is implemented in CFOUR program package \citep{cfour}, in conjunction with Dunning's correlation consistent triple-$\zeta$ (cc-pVTZ) basis set \citep{Dunning1989}.
The convergence criteria for the HF-SCF equations, the CC amplitude equations and
the linear equations were set to 10$^{-8}$ atomic units.
This means the convergence structure calculations and the analytical second derivatives were followed by finite difference techniques to obtain the full
cubic force field. These harmonic and anharmonic force field calculations yielded the rotational and centrifugal distortion constants listed in Table \ref{constants} together
with the energy difference between the \textit{cis} and \textit{trans} isomer of 3.6~kJ~mol$^{-1}$ (301~cm$^{-1}$ or 433~K).
The optimized geometries of both species are shown in Fig. \ref{spectrum} and their harmonic and anharmonic vibrational frequencies are provided in Table \ref{vib-modes}.

     \begin{figure}[ht]
   \centering
   \includegraphics[trim = 0mm 20mm 0mm 25mm, clip, width=8.5cm]{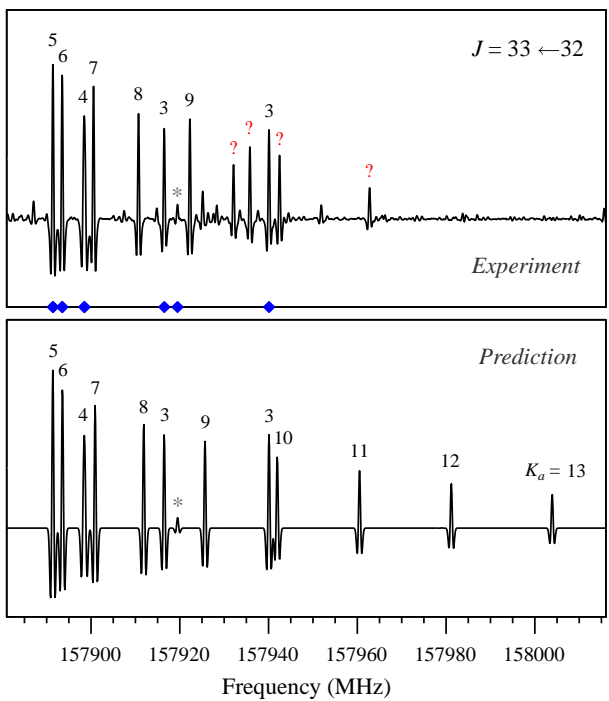}
      \caption{Perturbations in the ground state rotational spectrum of \textit{trans} vinyl isocyanate.
      The experimental spectrum is compared with predictions based on the analysis up to $K_a=6$.
      All the assigned lines correspond to $a$-type $R$-branch transitions except that marked with a star which is the $b$-type $40_{0,40}\leftarrow 39_{1,39}$ transition. The blue diamond symbols highlight the transitions included in the fit while the red question marks indicate perturbed transitions that could not be confidently assigned.}
      \label{perturbations}
   \end{figure}

\section{Rotational spectra and analyses}
{\label{s:analysis}}

\subsection{Trans isomer}

Since the \textit{trans} isomer represents the most stable species on the potential energy landscape of vinyl isocyanate, it is the most relevant
target for astronomical observations as it presents the strongest lines in the millimeter wave spectrum in Fig. \ref{spectrum}.
The most visible features in the spectrum are compact groups of $a$-type $R$-branch transitions arising from the near-prolate character of the molecule ($\kappa$ = --0.997) and its large dipole moment component along the $a$ principal inertial axis ($|\mu_{a}|=$ 2.047(6)~D and $|\mu_{b}| =$ 0.824(9)~D from \citealt{Kirby1978}).
The analysis was commenced by refitting the data set from \cite{Kirby1978} and generating spectral line predictions.
The predicted transitions were searched for with the help of the Loomis-Wood-type plot technique \citep{LW} implemented in our own program \citep{Vavra2020}.
The program is written in the Matlab software \citep{MATLAB:2020} and allows for the line assignments to quantum numbers, precise determination of the line frequencies employing the Voight profile function, and saving the data in a line list among other features. It generates input files for the SPFIT/SPCAT program package \citep{Pickett1991} which is executed directly from the user interface, allowing a straightforward analysis of the assigned lines. On this basis, we easily expanded the assignments for $K_{a}=$ 0--6 transitions up to $J = 69$ and identified weak $b$-type $R$-branch and $Q$-branch transitions.

Difficulties were encountered in assignments and fitting $K_{a}\geq 7$ transitions. In the absence of perturbations, these transitions are expected to progressively run to higher frequencies as shown in the lower panel of Fig. \ref{perturbations}. Instead, we observed irregular shifts with respect to their predicted positions that could not be treated within the scope of classical semi-rigid rotor Hamiltonian.
Some of these transitions even could not be confidently assigned (see Fig. \ref{perturbations}).
Consequently, we limited the analysis of $a$-type $R$-branch transitions to $K_{a}=6$.
These transitions were merged with $b$-type $R$-branch and $Q$-branch transitions which, due to low intensities and the large value of the $A$ rotational constant, were limited to $K_{a}=3$ and $J=80$.
Our data set was finally combined with microwave transitions from \cite{Kirby1978} and analyzed using Watson's $A$-reduced Hamiltonian in I$^{\text{r}}$-representation \citep{Watson1977} with terms up to the sixth power in the angular momentum. The only exceptions were $\Phi_K$ and $\phi_K$ which were fixed at the values estimated from quantum-chemical calculations. We note that all transitions from \cite{Kirby1978} were found to be fully consistent with our measurements and could be perfectly fitted. Results from this joint analysis are provided in Table \ref{constants} and the list of measured transitions in Table \ref{transitions-trans}. Watson’s $S$-reduced Hamiltonian led to results of similar quality.
The origin of perturbations in the rotational spectrum of \textit{trans} vinyl isocyanate is further discussed in Sect. \ref{ss:lab}.

\begin{table*}
\caption{Partition functions and abundances for the two isomers of vinyl isocyanate.}
\label{part-fce}
\begin{center}
\begin{footnotesize}
\setlength{\tabcolsep}{7.0pt}
\begin{tabular}{ r r r r r r r r r r r}
\hline\hline
\vspace{-0.3cm}\\
   &  \multicolumn{3}{c}{\textit{Trans} isomer} & & \multicolumn{3}{c}{\textit{Cis} isomer} & & & \\
\cline{2-4}
\cline{6-8}
\vspace{-0.3cm}\\
$T$ (K) &  $Q_{\text{rot}}$ & $Q_{\text{vib}}$ & (\%)\tablefootmark{a} & & $Q_{\text{rot}}$\tablefootmark{b} & $Q_{\text{vib}}$ & (\%) & & $Q_{\text{rot}}^{trans+cis}$ & $Q_{\text{vib}}^{trans+cis}$ \\
\vspace{-0.3cm}\\
\hline
\vspace{-0.3cm}\\
   300.000             &   46450.52  &  9.03          &  74          &  &  16049.82   &  9.22          &  26         & & 62500.34  &  9.08  \\
   225.000             &   30146.59  &  4.89          &  82          &  &   6435.09   &  5.00          &  18         & &  36581.68  &  4.91  \\
   150.000             &   16397.84  &  2.65          &  92          &  &   1336.05   &  2.69          &   8         & &  17733.89  &  2.65  \\
    75.000             &    5794.43           &  1.39          & 100          &  &     26.29   &  1.38          &   0         & &    5820.71  &  1.39  \\
    37.500             &    2048.88           &  1.06          & 100          &  &      0.03            &  1.05          &   0         & &     2048.91          &      1.06           \\
    18.750             &     724.98           &  1.00          & 100          &  &      0.00            &  1.00          &   0         & &     724.98           &      1.00           \\
     9.375             &     256.81           &  1.00          & 100          &  &      0.00            &  1.00          &   0         & &     256.81           &      1.00           \\
     5.000    &     100.37  &  1.00 & 100 &  &      0.00   &  1.00 &   0& &     100.37  &      1.00  \\
     2.725    &     40.65  &  1.00 & 100 &  &      0.00   &  1.00 &   0& &     40.65  &      1.00  \\
\hline
\end{tabular}
\end{footnotesize}
\end{center}
\tablefoot{
\tablefoottext{a}{Isomer abundance calculated as $Q_{\text{tot}}^{\mathit{trans}}/Q_{\text{tot}}^{\mathit{trans+cis}}$ where $Q_{\text{tot}}=Q_{\text{rot}}\times Q_{\text{vib}}$.}
\tablefoottext{b}{Corrected for the energy difference between the \textit{trans} and \textit{cis} isomer, i.e. the $0_{0,0}$ level is set to 301~cm$^{-1}$.}
}

\end{table*}

\begin{table*}[!ht]
 \begin{center}
 \caption{
 Parameters of our best-fit LTE model of methyl isocyanate toward Sgr~B2(N1S) and upper limits for ethyl isocyanate and vinyl isocyanate.
}
 \label{t:coldens}
 \vspace*{-1.2ex}
 \begin{tabular}{lcrccccccr}
 \hline\hline
 \multicolumn{1}{c}{Molecule} & \multicolumn{1}{c}{Status\tablefootmark{a}} & \multicolumn{1}{c}{$N_{\rm det}$\tablefootmark{b}} & \multicolumn{1}{c}{Size\tablefootmark{c}} & \multicolumn{1}{c}{$T_{\mathrm{rot}}$\tablefootmark{d}} & \multicolumn{1}{c}{$N$\tablefootmark{e}} & \multicolumn{1}{c}{$F_{\rm vib}$\tablefootmark{f}} & \multicolumn{1}{c}{$\Delta V$\tablefootmark{g}} & \multicolumn{1}{c}{$V_{\mathrm{off}}$\tablefootmark{h}} & \multicolumn{1}{c}{$\frac{N_{\rm ref}}{N}$\tablefootmark{i}} \\
  & & & \multicolumn{1}{c}{\small ($''$)} & \multicolumn{1}{c}{\small (K)} & \multicolumn{1}{c}{\small (cm$^{-2}$)} & & \multicolumn{1}{c}{\small (km~s$^{-1}$)} & \multicolumn{1}{c}{\small (km~s$^{-1}$)} & \\
 \hline
 CH$_3$NCO\tablefootmark{(j)}$^\star$ & d & 51 &  2.0 &  200 &  2.5 (17) & 1.00 & 5.0 & 0.0 &       1 \\
\hline
 C$_2$H$_5$NCO, $\varv=0$ & n & 0 &  2.0 &  200 & $<$  8.1 (16) & 10.1 & 5.0 & 0.0 & $>$     3.1 \\
 \hspace*{10ex} $\varv_{\rm t}=1$ & n & 0 &  2.0 &  200 & $<$  8.1 (16) & 10.1 & 5.0 & 0.0 & $>$     3.1 \\
\hline
 \textit{trans}-C$_2$H$_3$NCO, $\varv=0$ & n & 0 &  2.0 &  200 & $<$  2.4 (16) & 4.00 & 5.0 & 0.0 & $>$      11 \\
 \textit{cis}-C$_2$H$_3$NCO, $\varv=0$ & n & 0 &  2.0 &  200 & $<$  5.0 (17) & 4.00 & 5.0 & 0.0 & $>$    0.50 \\
\hline
 \end{tabular}
 \end{center}
 \vspace*{-2.5ex}
 \tablefoot{
 \tablefoottext{a}{d: detection, n: nondetection.}
 \tablefoottext{b}{Number of detected lines \citep[conservative estimate, see Sect.~3 of][]{Belloche16}. One line of a given species may mean a group of transitions of that species that are blended together.}
 \tablefoottext{c}{Source diameter (\textit{FWHM}).}
 \tablefoottext{d}{Rotational temperature.}
 \tablefoottext{e}{Total column density of the molecule. $x$ ($y$) means $x \times 10^y$.}
 \tablefoottext{f}{Correction factor that was applied to the column density to account for the contribution of vibrationally excited states, in the cases where this contribution was not included in the partition function of the spectroscopic predictions.}
 \tablefoottext{g}{Linewidth (\textit{FWHM}).}
 \tablefoottext{h}{Velocity offset with respect to the assumed systemic velocity of Sgr~B2(N1S), $V_{\mathrm{sys}} = 62$ km~s$^{-1}$.}
 \tablefoottext{i}{Column density ratio, with $N_{\rm ref}$ the column density of the previous reference species marked with a $\star$.}
 \tablefoottext{j}{The parameters were derived from the ReMoCA survey by \citet{Belloche19}.}
 }
 \end{table*}

\subsection{Cis isomer}

The spectroscopic constants from \cite{Kirby1978} were used for the first prediction of the rotational spectrum of \textit{cis} vinyl isocyanate in the millimeter wave region.
Only $a$-type $R$-branch transitions were observed for this isomer in agreement with the dipole moment components of $|\mu_{a}|=$ 2.14~(2)~D and $|\mu_{b}| =$~0.09~(2)~D as determined by Stark spectroscopy \citep{Bouchy1979}.
It was possible to identify and iteratively fit the rotational transitions up to $J=61$ and $K_a=21$.
We were not able to assign the transitions with higher values of $K_a$ because of their low intensities, which caused them to disappear in the spectral confusion,"weeds", of lines from the most stable \textit{trans} isomer.
Our data set was combined with low-frequency transitions from \cite{Kirby1978} and \cite{Bouchy1977} and was
fit to matrix elements of Watson's $A$-reduced effective rotational Hamiltonian \citep{Watson1977}. Some transitions from previous microwave works revealed larger residuals.
Larger uncertainties were thus assigned to these transitions and four of them were excluded from the fit.
The complete list of treated transitions is provided in Table \ref{transitions-cis}.
and the molecular constants determined from the analysis are reported in Table~\ref{constants}.

\begin{figure*}
\centerline{\resizebox{0.75\hsize}{!}{\includegraphics[angle=0]{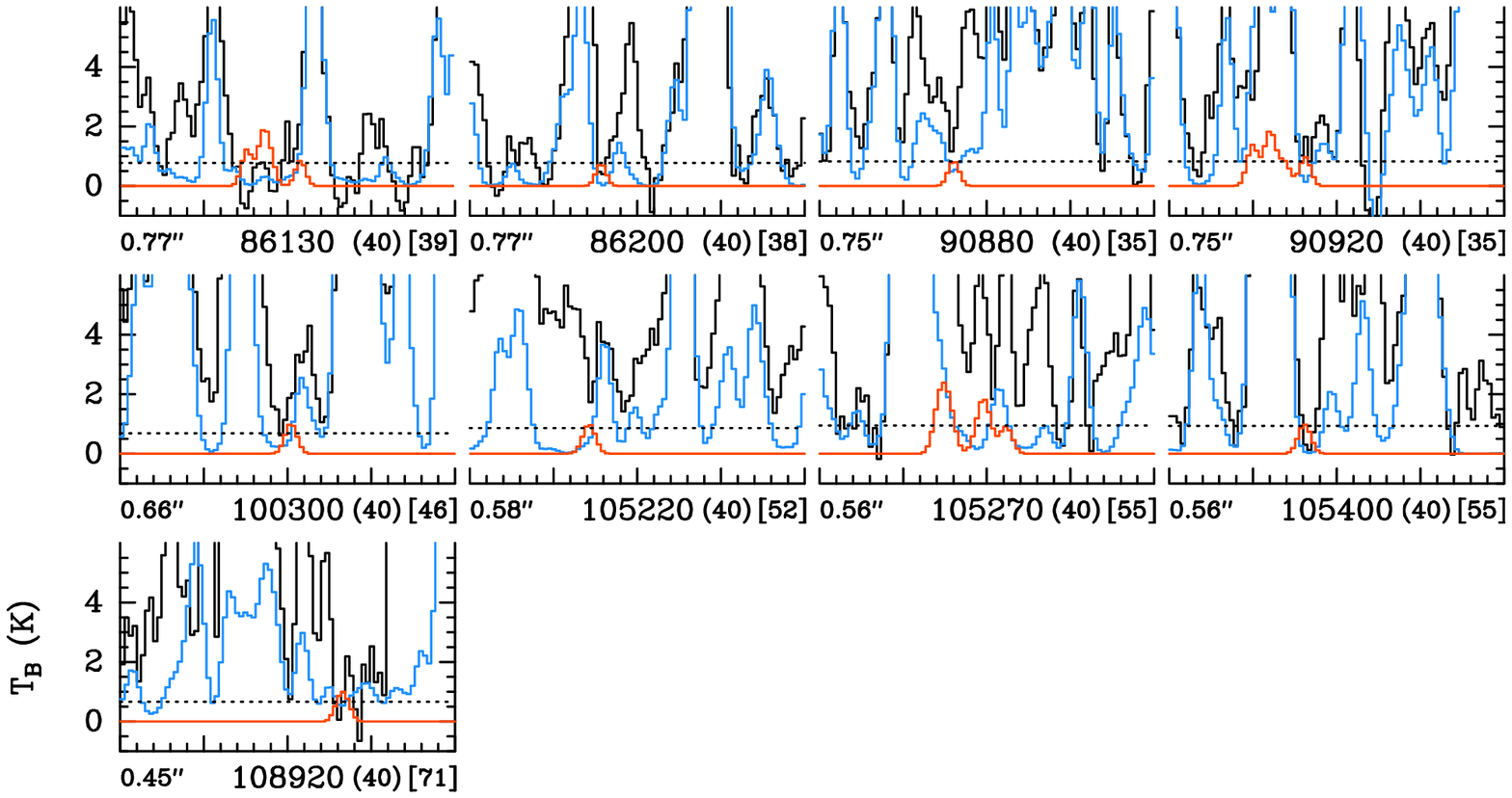}}}
\caption{Transitions of vinyl isocyanate \textit{trans}-C$_2$H$_3$NCO covered
by the ReMoCA survey. The LTE synthetic spectrum used to derive the upper limit
on the column density of \textit{trans}-C$_2$H$_3$NCO, $\varv = 0$ is
displayed in red and overlaid on the observed spectrum of Sgr~B2(N1S) shown in
black. The blue synthetic spectrum contains the contributions of all molecules
identified in our survey so far, but does not include the contribution of the
species shown in red. The central frequency is indicated in MHz below each
panel as well as the half-power beam width on the left, the width of each
panel in MHz in parentheses, and the continuum level of the
baseline-subtracted spectra in K in brackets. The y-axis is labeled in
brightness temperature units (K). The dotted line indicates the $3\sigma$
noise level. The figure only shows the transitions of \textit{trans}-C$_2$H$_3$NCO for which the red
synthetic spectrum has a significant peak temperature (compared to the noise
level) and which are not too heavily blended with much stronger emission of
other molecules.}
\label{f:spec_c2h3nco-t_ve0}
\end{figure*}

\subsection{Partition functions}

The spectroscopic constants from Table \ref{constants} were used to evaluate the rotational partition function ($Q_{\text{rot}}$) for both isomers.
We used the SPCAT program \citep{Pickett1991} to undertake the numerical summation over the ground state energy levels up to $J=240$ and $K_{a}=46$ for the \textit{trans} isomer and $J=210$ and $K_{a}=72$ for the \textit{cis} isomer. In addition, for the \textit{cis} form, this summation was corrected for the energy difference between the \textit{cis} and \textit{trans} form.
We used $\Delta E_{\text{ZPE}}=301$~cm$^{-1}$ from our quantum-chemical calculations. This value reproduces quite well the experimental intensities of the \textit{cis} isomer lines with respect to neighboring lines of the \textit{trans} isomer in our spectrum. On the other hand, the experimental value 417(20)~cm$^{-1}$ from \cite{Kirby1978} underestimates these \textit{cis} isomer lines relative intensities.
For this reason we consider our calculated value for the energy difference as more reliable.
Obtained rotational partition functions at nine different temperatures are provided in Table \ref{part-fce}. Their values represent the individual contributions from the \textit{trans} and \textit{cis} isomer to the rotational partition function of the molecule as a whole which is also given in Table \ref{part-fce}.
For completeness, we provide in Table \ref{part-fce-cisE0} the rotational partition functions of the \textit{cis} isomer also as a separate species, i.e. with the $0_{0,0}$ level set to 0~cm$^{-1}$.
The vibrational partition functions ($Q_{\text{vib}}$) were estimated using Eq. 3.60 of \cite{Gordy1970} by taking into account the anharmonic frequencies of the eighteen normal vibrational modes from Table~\ref{vib-modes} and are listed in Table \ref{part-fce}.
The same table also shows that the \textit{cis} isomer represents an important fraction of the room-temperature population of
vinyl isocyanate while its abundance is estimated to only 8~\% at 150~K.

\begin{figure*}
\centerline{\resizebox{0.75\hsize}{!}{\includegraphics[angle=0]{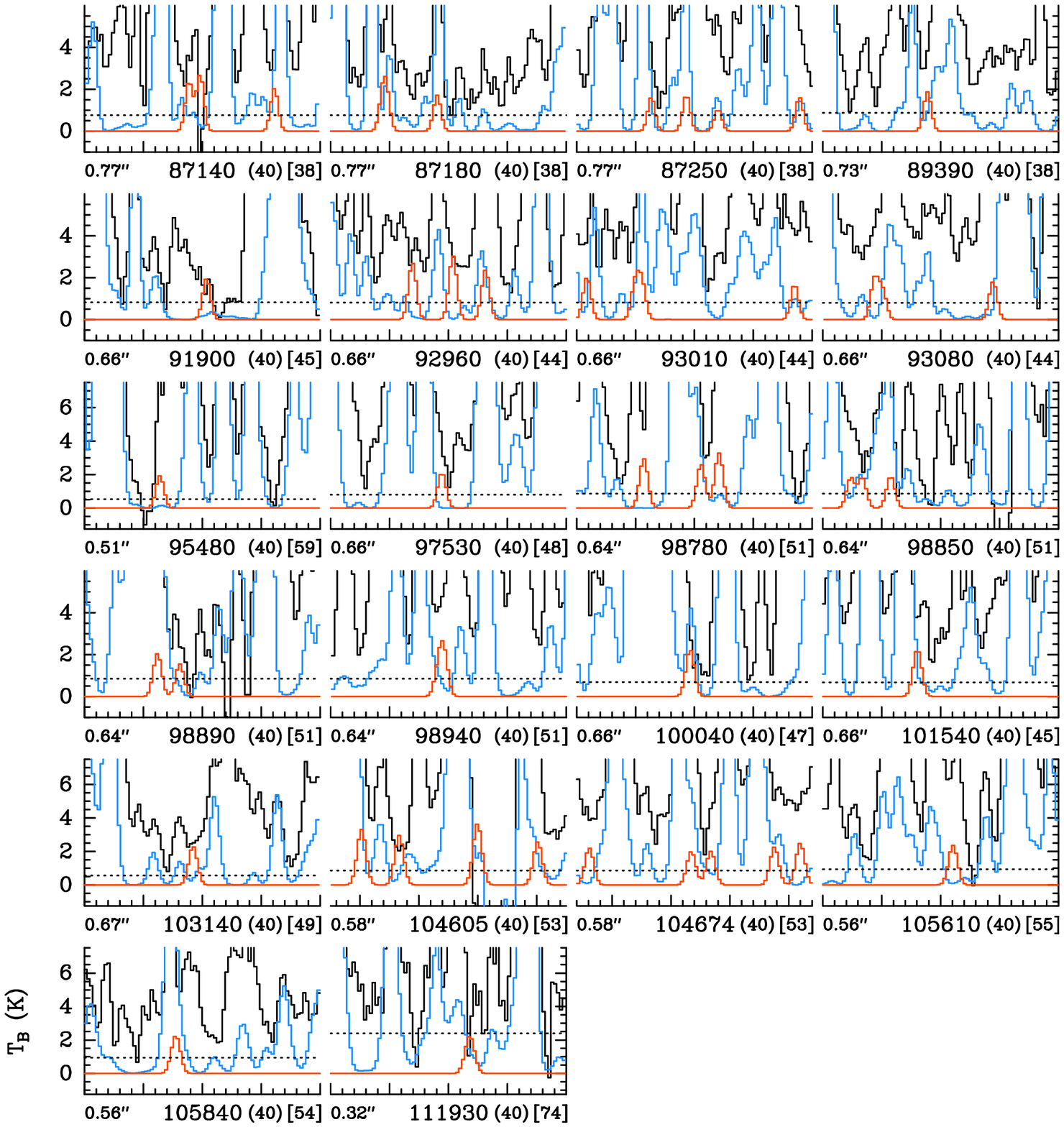}}}
\caption{Same as Fig.~\ref{f:spec_c2h3nco-t_ve0} but for
\textit{cis}-C$_2$H$_3$NCO, $\varv$~=~0.}
\label{f:spec_c2h3nco-c_ve0}
\end{figure*}

\begin{figure*}
\centerline{\resizebox{0.75\hsize}{!}{\includegraphics[angle=0]{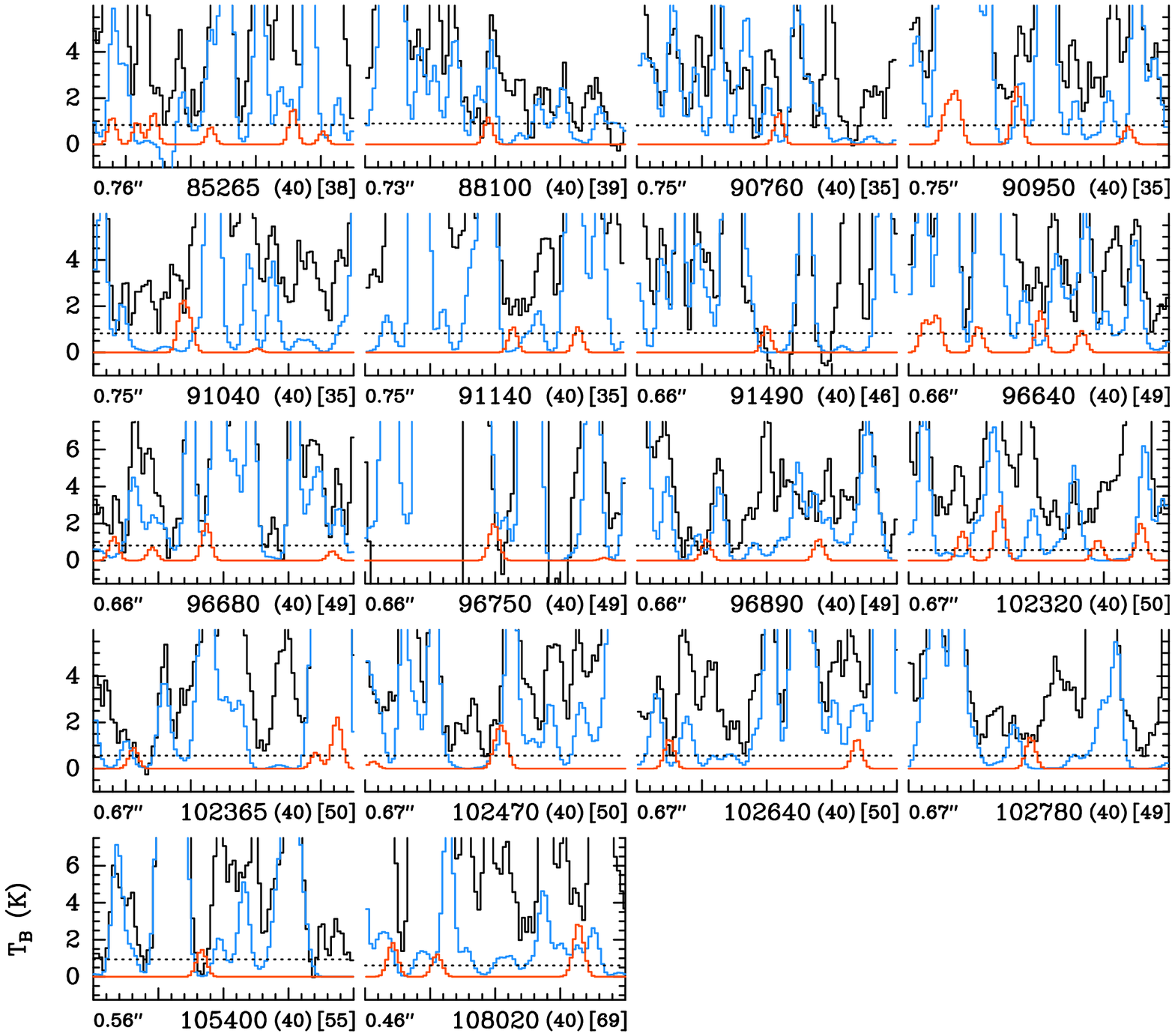}}}
\caption{Same as Fig.~\ref{f:spec_c2h3nco-t_ve0} but for C$_2$H$_5$NCO,
$\varv$~=~0.}
\label{f:spec_c2h5nco_ve0}
\end{figure*}

\section{Search for vinyl isocyanate toward Sgr~B2(N1)}
\label{s:astro}

\subsection{Observations}
\label{ss:obs_remoca}

The imaging spectral line survey Reexploring Molecular Complexity with ALMA
(ReMoCA) was performed toward the high-mass star forming protocluster
Sgr~B2(N) with the Atacama Large Millimeter/submillimeter Array (ALMA). A
detailed description of the observations and data reduction can be found in
\citet{Belloche19}. We summarize the main features of the survey here. The
phase center is located at the equatorial position
($\alpha, \delta$)$_{\rm J2000}$=
($17^{\rm h}47^{\rm m}19{\fs}87, -28^\circ22'16{\farcs}0$) which is halfway
between the two hot molecular cores Sgr~B2(N1) and Sgr~B2(N2). We defined five
frequency tunings to cover the frequency range from 84.1~GHz to
114.4~GHz with a spectral resolution of 488~kHz (1.7 to 1.3~km~s$^{-1}$).
The survey has a sensitivity per spectral channel that varies
between 0.35~mJy~beam$^{-1}$ and 1.1~mJy~beam$^{-1}$ (rms) depending on the
setup, with a median value of 0.8~mJy~beam$^{-1}$. The angular resolution
(HPBW) ranges from $\sim$0.3$\arcsec$ to $\sim$0.8$\arcsec$ with a median
value of 0.6$\arcsec$ that corresponds to $\sim$4900~au at the distance
of Sgr~B2 \citep[8.2~kpc,][]{Reid19}. An improved version of the
data reduction as described in \citet{Melosso20} was used for this work.
A detailed description of the procedure that was followed to subtract the
continuum emission can be found in Sect. 2.2 of \citet{Belloche19},
complemented by Sect. 4.1 of \citet{Melosso20}. It is difficult to estimate
the uncertainty on the subtracted continuum level in a robust way because of
spectral confusion. From our experience in modeling the line spectra of the
ReMoCA survey, we believe that this uncertainty may be in some cases on the
same order as the noise level itself. We emphasize however that our continuum
subtraction procedure is applied to each spectral window of 1.8~GHz width as a
whole, therefore it is not affected by spectral confusion that arises at
scales of tens or hundreds of MHz.

\begin{figure*}
\centerline{\resizebox{0.75\hsize}{!}{\includegraphics[angle=0]{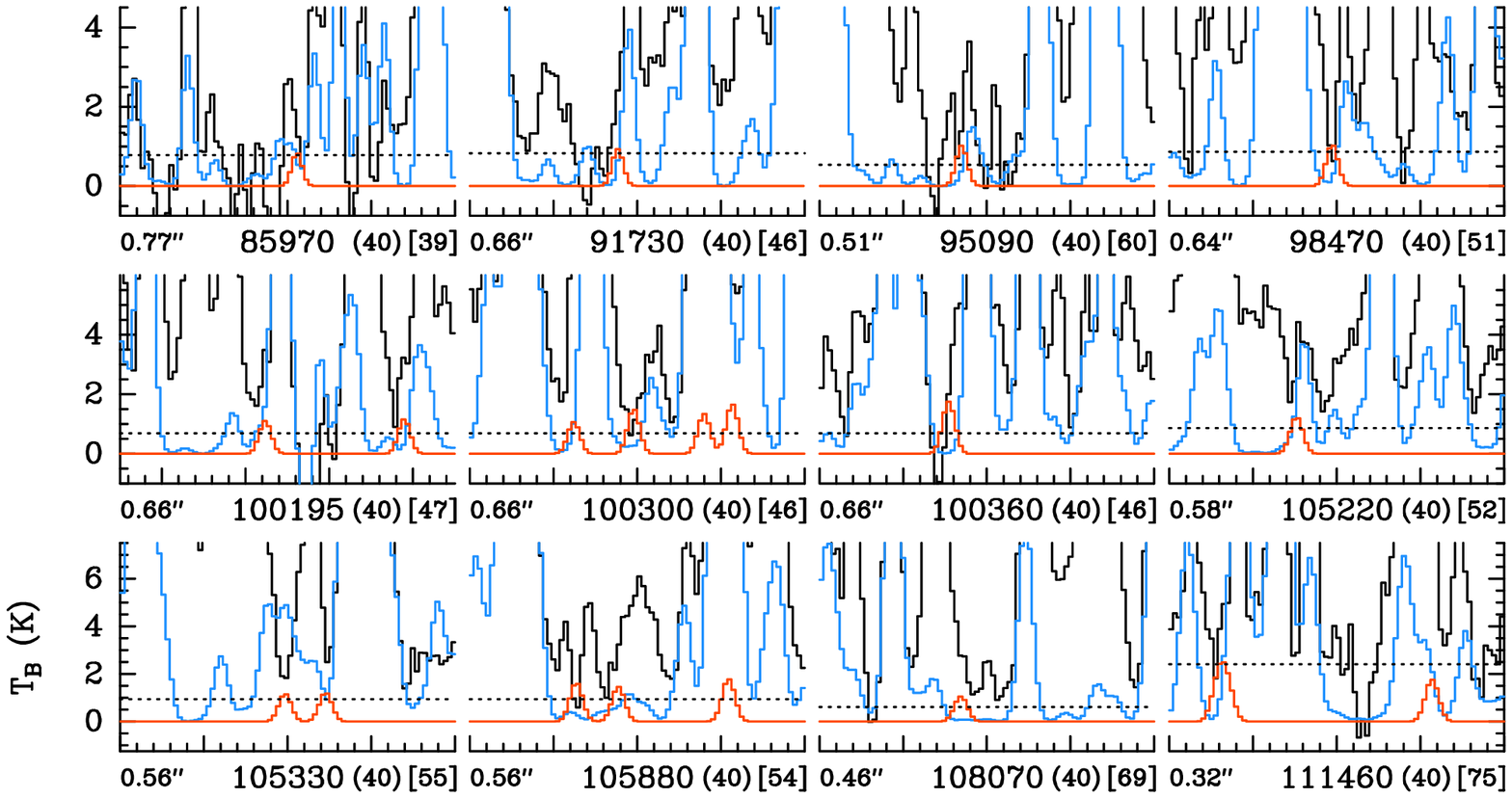}}}
\caption{Same as Fig.~\ref{f:spec_c2h3nco-t_ve0} but for C$_2$H$_5$NCO,
$\varv_{\rm t}$~=~1.}
\label{f:spec_c2h5nco_ve1}
\end{figure*}

We analyzed the spectrum obtained toward the position Sgr~B2(N1S) at
($\alpha, \delta$)$_{\rm J2000}$=
($17^{\rm h}47^{\rm m}19{\fs}870$, $-28^\circ22\arcmin19{\farcs}48$) following the
strategy employed by \citet{Belloche19}. This position is offset by about
1$\arcsec$ to the south of the main hot core Sgr~B2(N1) and has a lower
continuum opacity compared to the peak of the hot core. We assumed local
thermodynamic equilibrium (LTE) and produced synthetic spectra with the
astronomical software Weeds \citep[][]{Maret11} in order to analyze the
observed spectrum. The LTE assumption is justified by the high
densities of the regions where hot-core emission is detected in Sgr~B2(N)
\citep[$>1 \times 10^{7}$~cm$^{-3}$, see][]{Bonfand19}.
We derived a best-fit synthetic spectrum for each molecule separately, and
then added the contributions of all identified molecules together. We used a
set of five parameters to model the contribution of each species: size of the
emitting region
($\theta_{\rm s}$), column density ($N$), temperature ($T_{\rm rot}$), linewidth
($\Delta V$), and velocity offset ($V_{\rm off}$) with respect to the assumed
systemic velocity of the source, $V_{\rm sys}=62$~km~s$^{-1}$.
For nondetected species, the
synthetic spectra that were used to derive their column density upper limits
(red spectra in the figures) are conservative in the sense that they have
synthetic peak temperatures that are sometimes somewhat higher than the
3$\sigma$ noise level (dotted lines in the figures) or slightly above the
signals detected by ALMA (black spectra in the figures), implicitly accounting
for the additional (uncertain) uncertainty that affects the baseline level. In
this way, we are confident that the upper limits do not underestimate the
actual column densities of the nondetected species. For molecules that are detected (see, e.g., \citealt{Belloche19}), the
emission size is measured with Gaussian fits to the integrated intensity maps
of their uncontaminated transitions. The velocity offset and velocity width
are determined directly from the individual line profiles. The rotation
temperature is estimated from a population diagram. Finally, the only
remaining free parameter, the column density, is adjusted manually until a
good visual match between the synthetic and observed spectra is obtained. For
an undetected species, the first four parameters are fixed to values obtained
for a related species and the column density is varied until discrepancies at
the $\sim$3$\sigma$ level appear between the synthetic and observed spectra.
This yields the column density upper limit of the undetected species.

\subsection{Nondetection of vinyl isocyanate}
\label{ss:nondetection_remoca}

In order to search for vinyl isocyanate, C$_2$H$_3$NCO, toward Sgr~B2(N1S),
we relied on the LTE parameters derived for methyl isocyanate, CH$_3$NCO,
toward the same source by \citet{Belloche19} with the ReMoCA survey. These
parameters are listed in Table~\ref{t:coldens}. Assuming that the more complex
molecule vinyl isocyanate traces the same region as methyl isocyanate, we
produced LTE synthetic spectra for the former species adopting the same
parameters as for the latter with only the column density left as a free
parameter. We employed the spectroscopic predictions derived for the
\textit{trans} and \textit{cis} isomers of vinyl isocyanate in
Sect.~\ref{s:analysis} to compute their LTE synthetic spectra. None of these
conformers is detected toward Sgr~B2(N1S), as illustrated in
Figs.~\ref{f:spec_c2h3nco-t_ve0} and \ref{f:spec_c2h3nco-c_ve0}. The upper
limit on the total column density of vinyl isocyanate derived from each
conformer is reported in Table~\ref{t:coldens},
after accounting for the vibrational partition function $Q_{\rm
vib}^{trans+cis}$ provided in Sect.~\ref{s:analysis}.

We also report in Table~\ref{t:coldens} the column density upper limit that we
obtained with the ReMoCA survey for ethyl isocyanate, C$_2$H$_5$NCO. The
nondetection of this molecule, both in its torsional ground state and its
first torsionally excited state, is illustrated in
Figs.~\ref{f:spec_c2h5nco_ve0} and \ref{f:spec_c2h5nco_ve1}, respectively.
We employed the spectroscopic entry 71508 (version 1) of the Cologne Database
for Molecular Spectroscopy\footnote{https://cdms.astro.uni-koeln.de/}
\citep[CDMS,][]{Mueller05} to compute the LTE synthetic spectra of the
torsional ground state used to derive the upper limit to the column density of
ethyl isocyanate. This CDMS entry is mainly based on the measurements reported
in \citet{Kolesnikova18}. For the first torsionally excited state, we used
spectroscopic predictions from \citet{Kolesnikova18} prepared in electronic
format by one of us. The upper limit given in Table~\ref{t:coldens} accounts
for the (substantial) vibrational correction that was estimated using the
energies of the vibrational modes derived by \citet{Durig10} for ethyl
isocyanate. This upper limit holds under the assumption that a single conformation
exists for this species. A single form with quite a large number of molecules
in excited torsional states is suggested from the room-temperature microwave
and millimeter wave spectroscopy of gas phase samples \citep{Sakaizumi1976,Kolesnikova18}. Infrared spectroscopic data of
ethyl isocyanate dissolved in liquid noble gases were, on the other hand, interpreted
as a mixture of the \textit{cis} and \textit{trans} forms on the basis of quantum-chemical calculations \citep{Durig10}.
These calculations are, however, strongly dependent on the basis set used and
predict very low barrier for the conformational interchange which could even
fall into a calculation error.

The most stringent constraint on the column density of vinyl isocyanate is
obtained from its \textit{trans} conformer. We find that vinyl isocyanate is
at least 11 times less abundant than methyl isocyanate toward Sgr B2(N1S).
This is a factor $\sim$2 less stringent than the limit found on the basis of
the \textit{trans} conformer of vinyl isocyanate by \cite{Rodriguez-Almeida21}
toward G+0.693-0.027, a shocked region located close to Sgr B2(N)
(CH$_3$NCO/\textit{trans}-C$_2$H$_3$NCO > 26). For the \textit{cis} conformer, considered as an independent
species, they obtained a lower limit of 7 for CH$_3$NCO/\textit{cis}-C$_2$H$_3$NCO but this cannot
be compared directly to the ratio we report in Table \ref{t:coldens} because, for the high
densities of Sgr B2(N1S), we assume an LTE distribution of both conformers,
and not independent species.
Because of its large vibrational partition function, the
upper limit we obtained for ethyl isocyanate toward Sgr B2(N1S) is less stringent than for vinyl isocyanate: we find that ethyl
isocyanate is at least 3 times less abundant than methyl isocyanate. For
comparison, propanal, C$_2$H$_5$CHO, and ethylamine, C$_2$H$_5$NH$_2$, were
both found to be at least 5 times less abundant than acetaldehyde, CH$_3$CHO,
and methylamine, CH$_3$NH$_2$, respectively, toward Sgr~B2(N1S) with the ReMoCA
survey \citep[][]{SanzNovo22,Margules22}. Therefore, it is likely that ethyl
isocyanate is at least twice less abundant than the upper limit derived above, which would be in line with the ratio
CH$_3$NCO/C$_2$H$_5$NCO of 8 found by \cite{Rodriguez-Almeida21} toward
G+0.693-0.027.
In contrast, the upper limit obtained for vinyl isocyanate toward Sgr~B2(N1S) is a bit more
stringent than the one derived toward the same source for vinylamine,
C$_2$H$_3$NH$_2$, that was found to be at least 8 times less abundant than
methylamine \citep[][]{Margules22}.

\section{Discussion}
\label{s:discussion}

\subsection{Laboratory spectroscopy of vinyl isocyanate}
\label{ss:lab}

\begin{figure}[ht]
   \centering
   \includegraphics[trim = 0mm 5mm 0mm 0mm, clip, width=8.9cm]{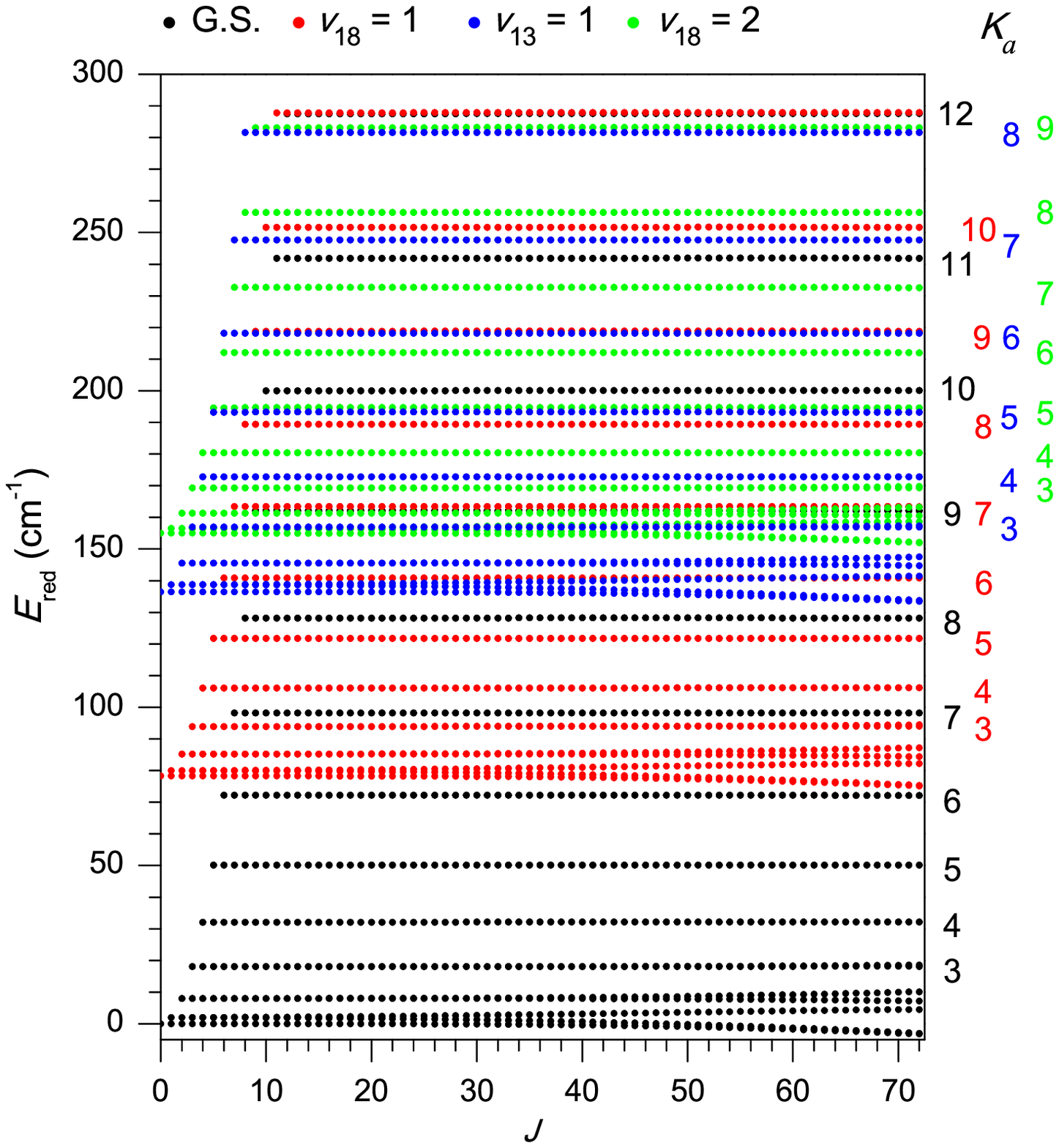}
      \caption{Diagram of reduced energies for the ground state (black), $v_{18}=1$ (red), $v_{13}=1$ (blue), and $v_{18}=2$ (green) excited vibrational states in \textit{trans} vinyl isocyanate. The reduced energies $E_{\text{red}}$ are calculated as $E_{\text{red}}=E - (B+C)J(J+1)/2$ where $E$ corresponds to the energy of rotational levels. Their values are obtained from the experimental spectroscopic constants and calculated vibrational energies.
      }
      \label{Ered}
   \end{figure}

Isocyanates are rather flexible molecules with plenty of anomalies in their rotational spectra \citep{Yamada1980,Koput1984,Cernicharo16,Pienkina2017,Kolesnikova18,Kolesnikova2019} and for vinyl isocyanate this has also proven to be the case. The ground state rotational spectrum of the \textit{trans} isomer suffered from strong perturbations which allowed only transitions involving levels up to $K_{a}=6$ to be analyzed without the loss of physical meaning of the fitted constants.
The origin of these perturbations can be understood once excited vibrational states are taken into consideration and the reduced energy level diagram (Fig. \ref{Ered}) is plotted.
Three excited vibrational states, namely $v_{18}=1$, $v_{13}=1$, and $v_{18}=2$, fall into the energy window of 200~cm$^{-1}$. Figure~\ref{spectrum} shows that our spectrum exhibits noticeable patterns assignable to these states on the basis of the spectroscopic constants from \cite{Kirby1978}, nevertheless, their low $K_{a}$ transitions were already heavily perturbed.
\cite{Kirby1978} estimated the frequencies of the associated vibrational modes $\nu_{18}$ and $\nu_{13}$ to 77(10)~cm$^{-1}$ and 200(20)~cm$^{-1}$, respectively, from microwave relative intensity measurements. We find a remarkable agreement between the literature value and the anharmonic frequency of 78~cm$^{-1}$ for the $\nu_{18}$ vibrational mode from our quantum-chemical calculations.
On the other hand, our computations indicated a significantly lower frequency for the $\nu_{13}$ mode (136~cm$^{-1}$) than previously reported.
This lower value leads to better agreement with our experimental spectrum; the rotational transitions in $v_{13}=1$ have on average slightly higher relative intensities than the same transitions in $v_{18}=2$ which is estimated to lie at 155~cm$^{-1}$.

The excited vibrational states $v_{18}=1$ and $v_{13}=1$ are in strong non-resonant Coriolis interaction, repelling each other, as evidenced by their $A$ rotational constants (54~601 and 70~532~MHz for $v_{18}=1$ and $v_{13}=1$, respectively, \citealt{Kirby1978}). The rotational energy levels in $v_{18}=1$ are thus pushed down. Since the ground state rotational constant $A$ is considerable ($\sim2$~cm$^{-1}$ or 62~586~MHz) in comparison with the energy of $v_{18}=1$, the rotational levels in the ground state quickly reach in energy the levels in $v_{18}=1$ and rovibrational interactions might appear already for $K_{a}=7$ of the ground state (see Fig.~\ref{Ered}). Resonant interactions between $v_{18}=1$ and $v_{13}=1$ (e.g., between $K_{a}=8$ and 5, $K_{a}=9$ and 6, etc.) will further complicate the situation.
Figure~\ref{Ered} also shows that the $v_{13}=1$ state lies close in energy to $v_{18}=2$ and their rotational energy levels cross at $K_{a}=5$ and 6. Effects of this Fermi interaction
might be non-negligible as shown, for example, in \textit{n}-propyl cyanide \citep{Liu2019}.

All in all, the ground vibrational state cannot be completely analyzed without $v_{18}=1$ which cannot be treated without $v_{13}=1$ which in turn cannot be analyzed without $v_{18}=2$. Similar interactions among multiple vibrational states were observed in the rotational spectra of quasi-linear molecules such as HNCO \citep{Yamada1980} and hydrazoic acid (HN$_3$, \citealt{Hegelund1987,Vavra2017}) and other near-prolate species such as vinyl cyanide (C$_2$H$_3$CN, \citealt{Kisiel2009,Kisiel2012}) and \textit{n}-propyl cyanide (\textit{n}-C$_3$H$_7$CN, \citealt{Liu2019}).
The above network of interacting states is already very complicated and is expected to be even more complex due to possible coupling with other excited vibrational states which are not included in Fig.~\ref{Ered} for simplicity. In particular, $v_{18}=3$, ($v_{18}=1$, $v_{13}=1$), and $v_{13}=2$ might be at play and would make the analysis extremely challenging. A logical step toward the understanding of these interactions would be the measurement and analysis
of a high-resolution vibrational spectrum of vinyl isocyanate. For the time being, we prefer to make our laboratory data available for astrophysical applications even though our data set is limited in terms of $K_{a}$ in comparison with "well-behaved" molecular systems.
We emphasize that given the large value of the $A$ rotational constant, $K_{a}=6$ is more than enough to identify the molecule in fairly warm environments such as hot cores.
Of course, our results are also perfectly suitable for a search of the molecule in colder interstellar sources.
For CH$_3$NCO, only $K_{a}\leq 3$ transitions were experimentally accessible at the time of its search in space and led to its detection \citep{Halfen15,Cernicharo16}.

The set of the spectroscopic constants for \textit{trans} vinyl isocyanate obtained in this work is definitely more complete and accurate in comparison with the original work of \cite{Kirby1978}.
Table~\ref{constants} illustrates that the previous values for the rotational and quartic centrifugal distortion constants are improved by up to two orders of magnitude. Furthermore, the physical meaning of those newly determined quantities is corroborated by the agreement between the fitted values and their quantum-chemical counterparts in Table~\ref{constants}.
Spectral predictions for $K_{a} < 7$ generated using these constants thus serve as
an accurate observational reference, at least in the case when these are interpolations within the measured data sets. Although some transitions may reveal satisfactory predictive power outside the present data region, we do not recommend such extrapolations due to the amount and complexity of perturbations expected to appear at higher frequencies.

Final remarks concern the spectroscopic constants for \textit{cis} vinyl isocyanate.
We can infer from Table~\ref{constants} that their values are significantly better determined than previously; some of them by two orders of magnitude.
In addition, many centrifugal distortion constants are determined for the first time.
Table~\ref{constants} further illustrates that the \textit{cis} isomer analysis called for inclusion of centrifugal distortion terms up to 10$^{\text{th}}$ power in the angular momentum. The requirement for such a high expansion of the rotational Hamiltonian might be an indicative of rovibrational interactions with low-lying excited vibrational states. Here, the low-energy states $v_{18}=1$ and $v_{13}=1$ are predicted by our quantum-chemical calculations at 90 and 110 ~cm$^{-1}$, respectively. Since the rotational constant $A$ is three times smaller than that of the \textit{trans} isomer, such interaction would affect the rotational energy levels with higher $K_{a}$. To assess whether the interaction results in contributions to the centrifugal distortion constants, one can compare their experimental and quantum-chemical values.
Table \ref{constants} shows good correspondence between our quartic and sextic constants and those calculated at CCSD/cc-pVTZ level of theory except for $\phi_{JK}$ which is opposite in sign.
Thus, if there existed an interaction with excited vibrational states it would probably become noticeable at the sextic and higher-order level of the Hamiltonian. Unfortunately, we did not find the rotational transitions in excited vibrational states in our spectra that could address this issue.

\subsection{The chemistry of methyl, vinyl, and ethyl isocyanate}
\label{s:chemistry}

\subsubsection{Methyl isocyanate}

While methyl isocyanate is the only one of the three molecules listed in
Table~\ref{t:coldens} to have been definitively detected in our survey,
the chemistry by which it forms is still uncertain.
\citet{Halfen15} proposed gas-phase reactions between the methyl radical,
CH$_3$, and either HNCO (isocyanic acid) or HOCN (cyanic acid) as plausible
mechanisms:
\begin{equation}
{\rm CH}_3 + {\rm HNCO} \rightarrow {\rm CH}_3{\rm NCO} + {\rm H} \label{reac1}
\end{equation}
\begin{equation}
{\rm CH}_3 + {\rm HOCN} \rightarrow {\rm CH}_3{\rm NCO} + {\rm H} \label{reac2}
\end{equation}
While HNCO is substantially more abundant than HOCN in the ISM
\citep[][]{Bruenken10}, it is also the more stable structure; as noted by
\citet{Halfen15}, reaction \ref{reac1} should be endothermic. However, a
precise determination of this is challenging, as the enthalpy of formation of
methyl isocyanate is poorly defined in the literature. The CRC Handbook of
Chemistry and Physics \citep[][]{Haynes17} provides only a value for the
liquid phase ($\Delta_{f}H^{0}(l)=-92.0$~kJ~mol$^{-1}$) rather than the gas
phase. But even crudely adjusting this value to take account of the enthalpy
of vaporization, it appears almost certain that reaction \ref{reac1} would be
highly endothermic (by more than 100~kJ~mol$^{-1}$).
Calculations by \citet{Majumdar2017} indeed indicate the reaction to be endothermic by
77~kJ~mol$^{-1}$, with a transition state that lies 83~kJ~mol$^{-1}$ above the entrance level.
Furthermore, it is likely
that even reaction \ref{reac2} is somewhat endothermic. Substantial activation
energy barriers might also be expected for both reactions, suggesting that
either route would be quite inefficient in the ISM.

\citet{Halfen15} also proposed ion-molecule reactions between HNCO/HOCN and
CH$_5^+$ to produce protonated methyl isocyanate; however, such processes
involving CH$_5^+$ typically result in proton transfer in laboratory
experiments, indicating that protonated HNCO/HOCN would be the preferred
products. It therefore remains unclear whether there is a plausible gas-phase
mechanism that could produce CH$_3$NCO in sufficient quantities to explain
observations. As noted by \citet{Cernicharo16}, the detection of CH$_3$NCO
only in high-temperature regions may indicate that methyl isocyanate
originates on dust-grain surfaces.

Chemical models have been used to study the formation of CH$_3$NCO and related
species on interstellar dust grains. \citet{Belloche17} considered a direct
radical-radical association reaction:
\begin{equation}
{\rm CH}_3 + {\rm OCN} \rightarrow {\rm CH}_3{\rm NCO} \label{reac3}
\end{equation}
The radicals would be produced by the photodissociation of, or
chemical H-abstraction from, CH$_4$ and HNCO in the dust-grain ice mantles,
while CH$_3$ could also be produced by repetitive H addition to atomic carbon
on the grain/ice surfaces. Experiments presented by \citet{Ligterink17} indeed
show that CH$_3$NCO may be formed by UV irradiation of mixed CH$_4$:HNCO ices.
The precise mechanism of production was not determined, but was assumed to be
either reaction \ref{reac3} or reaction \ref{reac1}, which would be
endothermic also in the solid phase. As with reaction \ref{reac1}, the
alternative branch CH$_3$ + HNCO $\rightarrow$ CH$_4$ + OCN is endothermic.
However, the production of excited CH$_3$ as a photoproduct of CH$_4$ might be
sufficient to remove these difficulties in both cases.
\citeauthor{Ligterink17} also proposed that the OCN$^-$ ion, which was
produced abundantly in the experiments, could be active in forming CH$_3$NCO.
In their observational study, \citet{Ligterink2021} concluded that reaction \ref{reac3}, occurring during the dark cloud stage of evolution, would be a plausible explanation for the uniform CH$_3$NCO/HNCO ratio of around 10\% observed toward various sources.

The astrochemical model of \citet{Belloche17} indicated that, given an
appropriate degree of grain-surface formation of HNCO, methyl isocyanate could
also be formed on the grains during the warm-up phase of a hot core, via
reaction \ref{reac3}. The radicals would become mobile, and thus reactive, at
elevated temperatures; the model results provided an acceptable match to the
observed abundances.

However, the more recent models of \citet{Garrod22} have challenged the purely
diffusive picture of hot-core grain-surface chemistry, by including so-called
non-diffusive reactions on the grain surfaces and within the bulk ices
\citep[][]{Jin20}. Furthermore, these models restrict the mobility of most
chemical species in the bulk, allowing only H and H$_2$ to diffuse internally.
Larger species may diffuse only on the grain/ice surfaces themselves. The
non-diffusive reaction mechanisms in the model nevertheless allow
radical-radical reactions in the bulk ice to occur, but with rates driven by
the processes that initiate the radical production that precedes them. In this
scenario, the efficiency of bulk-ice chemistry is only weakly determined by
the ice temperature, while radical reactions in the bulk can occur in
principle at any temperature if, for example, photodissociation of a stable
molecule in the bulk ice occurs in the presence of a radical with which one of
the photoproducts may rapidly react.

Unfortunately, in these new models, CH$_3$NCO is severely underproduced on the
grains; OCN radicals tend to recombine quickly with mobile H atoms, both on
the surface and in the bulk ice, providing little opportunity for reaction
with CH$_3$. Furthermore, the endothermicity of the H-abstraction reaction
H + HNCO $\rightarrow$ OCN + H$_2$ removes an alternative means of reforming
OCN and thus raising its abundance on the grains. Meanwhile, a competing
process, H + HNCO $\rightarrow$ NH$_2$CO, is exothermic with only a modest
barrier \citep[1390 K;][]{Nguyen96}. The removal of the bulk diffusion
mechanisms for OCN and CH$_3$ also inhibits the diffusive reactions that
produced CH$_3$NCO effectively at higher temperatures in the older models.
HNCO, however, is itself formed effectively on the grain surfaces at very low
temperatures in these models \citep[see][]{Garrod22}, through the
barrier-mediated reaction NH + CO $\rightarrow$ HNCO.

Thus, while photodissociation of CH$_4$ and HNCO, followed by radical addition,
still appears to be the main production mechanism for methyl isocyanate in the
newest astrochemical models, the resulting abundances are several orders of
magnitude lower than the observed values.

Several other gas-grain modeling studies have been conducted to explain observed gas-phase CH$_3$NCO abundances toward cold sources and warm star-forming regions \citep{Martin-Domenech2017,Quenard2018}. Those models adopted the gas-phase and grain-surface reaction mechanisms noted above; they concluded that reactions \ref{reac1} and \ref{reac2} are dominant contributors to CH$_3$NCO production following ice mantle desorption, based on the reactions occurring at the collisional rate, i.e. somewhere on the order of 10$^{-10}$ cm$^{3}$~s$^{-1}$. However, as noted above, the substantial endothermicities of those reactions would likely render them highly inefficient; some alternative formation mechanism would therefore still be required, whether in the gas phase or on the grains.

Noting the problems with the suggested gas-phase reactions, \citet{Majumdar2017} proposed two new mechanisms for grain-surface CH$_3$NCO production. The first involves the reaction of an H atom with an HCN molecule while the latter is in a van der Waals complex with a CO molecule; i.e.
\begin{equation}
{\rm H} + {\rm HCN}...{\rm CO} \rightarrow {\rm H}_2{\rm CN}...{\rm CO} \rightarrow {\rm CH}_2{\rm NCO} \label{reac_maj1}
\end{equation}
The energy produced in the initial reaction would allow the activation energy barrier to be overcome for an immediate follow-on reaction, producing CH$_2$NCO; the latter radical could then be rapidly hydrogenated by H to form methyl isocyanate. These authors' other mechanism involves a barrieless reaction between atomic N and the radical CH$_3$CO, for which they calculate that the production of CH$_3$NCO is the most energetically favorable outcome:
\begin{equation}
{\rm N} + {\rm CH}_3{\rm CO} \rightarrow {\rm CH}_3{\rm NCO} \label{reac_maj2}
\end{equation}

\citet{Majumdar2017} find that reactions \ref{reac_maj1} and \ref{reac_maj2} contribute to the production CH$_3$NCO sufficiently to explain observational values. However, the efficiency of reaction \ref{reac_maj1} is not well defined and is dependent on the precise reaction dynamics involved, as well as on the choice of barrier to the initial reaction. Reaction \ref{reac_maj2} relies on the hydrogenation of ketene (CH$_2$CO) to form the necessary radical, which is also a barrier-dependent process. The results are encouraging, although further testing of these mechanisms is likely required to determine their true effectiveness.

One further mechanism might be plausible for grain-surface production of
CH$_3$NCO. \citet{Garrod22} introduced a number of new reactions into their
chemical network involving methylene, CH$_2$. In its ground (triplet) state,
methylene is a diradical; \citeauthor{Garrod22} proposed that H-abstraction
reactions involving CH$_2$ and stable molecules on grain surfaces would lead
to production of two radicals that could immediately react with each other
to form a single product. It is unclear whether the initial abstraction
reaction between CH$_2$ and HNCO would be exothermic, but it would
almost certainly have an activation energy barrier.
However, a direct addition reaction,
\begin{equation}
{\rm CH}_2 + {\rm HNCO} \rightarrow {\rm CH}_3{\rm NCO} \label{reac4}
\end{equation}
would be strongly exothermic ($\sim$300~kJ~mol$^{-1}$). The attack
of the methylene radical on the C=N double bond in HNCO would likely have an
activation energy barrier, although the large exothermicity of the reaction
might suggest a relatively small barrier compared with, for example, the
above-mentioned addition reaction H + HNCO $\rightarrow$ NH$_2$CO (exothermic
by around 140~kJ~mol$^{-1}$). However, as the methylene reaction does not
explicitly involve an H atom, there may be no quantum tunneling mechanism
available to overcome the barrier, unlike in the case of many grain-surface
reactions involving atomic H. The determination of the efficiency of reaction
\ref{reac4} would require detailed calculations. But if CH$_2$ is a common
photodissociation product of CH$_4$, as is often assumed in the chemical
networks, then the spontaneous production of methylene in proximity to HNCO in
the bulk ice could plausibly produce methyl isocyanate in appreciable
quantities. Likewise, if the abstraction process is exothermic, then H-atom
tunneling between HNCO and CH$_2$ might allow it to proceed effectively,
producing CH$_3$NCO in a two-step process. As well as the bulk ice process,
formation of CH$_2$ on the grain/ice surface, from atomic carbon, might also
lead to reaction with HNCO.
Such a mechanism would presumably retain the observationally determined
correlation between HNCO and CH$_3$NCO \citep{Ligterink2021,Colzi2021}, once both molecules were desorbed into the gas phase at elevated temperatures.

Alternatively, as mentioned above, if CH$_3$ is formed in an excited state
through CH$_4$ photodissociation in the bulk ice, direct reaction
 of the methyl radical
with abundant
HNCO might proceed efficiently.

\subsubsection{Vinyl and ethyl isocyanate}

In spite of the nondetection of vinyl and ethyl isocyanate toward Sgr B2(N1S), it is valuable to
consider their possible production mechanisms. Although we do not currently
have chemical models that include either species, we may speculate on possible
outcomes based on molecules that are presently included.

The lack of a gas-phase formation mechanism for CH$_3$NCO would also suggest
that chemistry on grains or within the ice mantles is the most plausible
scenario for the larger molecules. We may again propose radical addition
reactions involving OCN, i.e.
\begin{equation}
{\rm C}_2{\rm H}_3 + {\rm OCN} \rightarrow {\rm C}_2{\rm H}_3{\rm NCO} \label{reac5}
\end{equation}
\begin{equation}
{\rm C}_2{\rm H}_5 + {\rm OCN} \rightarrow {\rm C}_2{\rm H}_5{\rm NCO} \label{reac6}
\end{equation}
Reaction \ref{reac6} was first suggested by \citet{Rodriguez-Almeida21}.
Assuming that reaction \ref{reac3} is the most important process for
CH$_3$NCO production, the relative abundances of the radicals CH$_3$,
C$_2$H$_3$, and C$_2$H$_5$ in the ice might indicate the expected ratio of the
isocyanates (on the expectation that the photodissociation of HNCO would
provide the driving rate for all three reactions, \ref{reac3}, \ref{reac5},
and \ref{reac6}). However, in the new \citet{Garrod22} models, the ratios of
the peak abundances (during the cold collapse stage) are approximately
CH$_3$:C$_2$H$_3$:C$_2$H$_5$ = 30:1:65. While this appears consistent
with the observed ratio of the column density of methyl isocyanate to the
upper limit for vinyl isocyanate, it would also indicate that the abundance of
ethyl isocyanate should exceed that of methyl isocyanate. As noted in
Sect.~\ref{ss:nondetection_remoca}, such a large ratio is not expected.

However, reaction \ref{reac3} does not produce adequate quantities of CH$_3$NCO
in the models. If an alternative process is active at least for the production
of this molecule, while reactions \ref{reac5} and \ref{reac6} were to remain
the principal routes to vinyl and ethyl cyanide, then this would indicate that
the real column densities of the latter species should be far less than the
observational upper limits.

If the methylene reaction proposed above -- reaction \ref{reac4} -- should in
fact be the dominant production route for CH$_3$NCO, then the absence of
analogous mechanisms that form vinyl and ethyl cyanide is readily understood;
there are no equivalent chemical species that would produce C$_2$H$_3$NCO or
C$_2$H$_5$NCO by a similar reaction with HNCO. In this case, the most likely
production mechanism for ethyl isocyanate could be the photodissociaton
of, or H-abstraction from, methyl isocyanate itself, followed by the addition of
a methyl group, e.g.:
\begin{equation}
{\rm CH}_3{\rm NCO} + h\nu \rightarrow {\rm CH}_2{\rm NCO} + {\rm H}
\end{equation}
\begin{equation}
{\rm CH}_3 + {\rm CH}_2{\rm NCO} \rightarrow {\rm C}_2{\rm H}_5{\rm NCO}
\end{equation}
The H-abstraction initiated process might include the two-step insertion of
methylene into a C-H bond in methyl isocyanate. The astrochemical models
indicate that in cases in which H-abstraction/photodissociation and methyl
addition is the main formation mechanism, such as with ethylamine production
from methylamine,
the smaller homologue achieves an abundance around 10 times greater than that of the larger homologue.
Such would be in line with our observational ratio of
methyl to ethyl
isocyanate. Furthermore, \citet{Rodriguez-Almeida21} recently detected
ethyl isocyanate toward the Galactic Center cloud G+0.693-0.027, finding a
ratio CH$_3$NCO:C$_2$H$_5$NCO = $8 \pm 1$, which is also in good agreement
with the typical modeling outcome.

Once again, there is unlikely to be a comparable mechanism leading to
formation of a vinyl group, which would also tend to make the abundance of
vinyl isocyanate inferior to those of the other two.

The above ideas are, of course, highly speculative. A full chemical model
including all of these species would be desirable, although there remain a
number of poorly-defined quantities that could be important -- not least, the
possible efficiency of reaction \ref{reac4}, and the precise mechanisms
involved in experimental UV-induced production of methyl isocyanate.

\section{Conclusions}
\textbf{\label{s:conclusions}}

Laboratory rotational spectroscopy of vinyl isocyanate has been undertaken in the frequency regions 127.5--218 and 285--330~GHz. Over 1000 transition lines were assigned and measured
for the ground vibrational states of its \textit{trans} and \textit{cis} isomers. The present work provides significantly more precise values of the spectroscopic parameters which agree with those from accompanying high-level quantum-chemical computations. We report a nondetection of both vinyl and ethyl isocyanate toward
the main hot core of Sgr~B2(N) that was targeted with ALMA. We find that these
molecules are at least 11 and 3 times less abundant than methyl isocyanate,
respectively.
Despite the nondetection of vinyl isocyanate in Sgr~B2(N) the present work represents a substantial improvement on previous microwave studies below 40 GHz and meets the requirements
for further searches of this species in the interstellar space.
To this end, spectral predictions are provided in Tables~\ref{predictions-trans} and \ref{predictions-cis} and will be also available in CDMS.

\begin{acknowledgements}

The spectroscopic part of this work has been funded by the Czech Science Foundation (GACR, grant No. 19-25116Y). L.K., K.V., J.K., and K.L. gratefully acknowledge this financial support.
L.K., J.K., K.L., and P.K. thank the financial support from the Ministry of Education, Youth and Sports of the Czech Republic (MSMT) within the Mobility grant No. 8J21FR006.
Computational resources were supplied by the project "e-Infrastruktura CZ" (e-INFRA CZ LM2018140) supported by the Ministry of Education, Youth and Sports of the Czech Republic. Computational resources were provided by the ELIXIR-CZ project (LM2018131), part of the international ELIXIR infrastructure.
R.T.G. thanks E. Herbst for helpful discussions.
This paper makes use of the following ALMA data:
ADS/JAO.ALMA\#2016.1.00074.S.
ALMA is a partnership of ESO (representing its member states), NSF (USA), and
NINS (Japan), together with NRC (Canada), NSC and ASIAA (Taiwan), and KASI
(Republic of Korea), in cooperation with the Republic of Chile. The Joint ALMA
Observatory is operated by ESO, AUI/NRAO, and NAOJ. The interferometric data
are available in the ALMA archive at https://almascience.eso.org/aq/.
Part of this work has been carried out within the Collaborative
Research Centre 956, sub-project B3, funded by the Deutsche
Forschungsgemeinschaft (DFG) -- project ID 184018867.
R.T.G. acknowledges funding from the Astronomy \& Astrophysics program
of the National Science Foundation (grant No. AST 19-06489).
J.-C.G. thanks the Barrande project No. 46662VH, the Centre National d'Etudes Spatiales (CNES) and the "Programme National Physique et Chimie du Milieu Interstellaire" (PCMI) of CNRS/INSU with INC/INP co-funded by CEA and CNES for a grant.

\end{acknowledgements}



\bibliography{library}



\begin{appendix}

\section{Complementary Tables}

Table \ref{vib-modes} lists the frequencies of normal vibrational modes of \textit{trans} and \textit{cis} isomers of vinyl isocyanate.
Table \ref{transitions-trans} lists the measured transitions of \textit{trans} vinyl isocyanate.
Table \ref{transitions-cis} lists the measured transitions of \textit{cis} vinyl isocyanate.
Table \ref{part-fce-cisE0} lists the rotational partition functions for \textit{cis} vinyl isocyanate with the $0_{0,0}$ energy level set to 0~cm$^{-1}$.
Table \ref{predictions-trans} lists the JPL/CDMS line catalog for \textit{trans} vinyl isocyanate.
Table \ref{predictions-cis} lists the JPL/CDMS line catalog for \textit{cis} vinyl isocyanate.

\begin{table*}
\caption{Harmonic and anharmonic frequencies of normal vibrational modes for \textit{trans} and \textit{cis} isomers of vinyl isocyanate calculated at CCSD/cc-pVTZ level of theory.}
\label{vib-modes}
\begin{center}
\begin{footnotesize}
\setlength{\tabcolsep}{7pt}
\begin{tabular}{r c c c r c c c }
\hline\hline
\vspace{-0.3cm}\\
     & \multicolumn{3}{c}{\textit{Trans} isomer} & &  \multicolumn{3}{c}{\textit{Cis} isomer} \\
\cline{2-4}
\cline{6-8}
\vspace{-0.3cm}\\
     & Harmonic  & Anharmonic  &   & & Harmonic  & Anharmonic  &   \\
Mode & frequency (cm$^{-1}$)   & frequency (cm$^{-1}$)  & Symmetry & & frequency (cm$^{-1}$) & frequency (cm$^{-1}$) & Symmetry \\
\vspace{-0.3cm}\\
\hline
\vspace{-0.3cm}\\
 1    & 3291.1  &  3149.9   &  $A^\prime$                          &   & 3281.7  & 3140.2   & $A^\prime$            \\
 2    & 3210.0  &  3079.0   &  $A^\prime$                          &   & 3239.1  & 3109.8   & $A^\prime$            \\
 3    & 3190.9  &  3051.5   &  $A^\prime$                          &   & 3183.8  & 3045.1   & $A^\prime$            \\
 4    & 2363.5  &  2301.7   &  $A^\prime$                          &   & 2359.8  & 2304.7   & $A^\prime$            \\
 5    & 1729.6  &  1635.6   &  $A^\prime$                          &   & 1713.5  & 1663.6   & $A^\prime$            \\
 6    & 1519.4  &  1490.3   &  $A^\prime$                          &   & 1531.5  & 1500.8   & $A^\prime$            \\
 7    & 1429.9  &  1408.3   &  $A^\prime$                          &   & 1452.3  & 1419.2   & $A^\prime$            \\
 8    & 1355.0  &  1328.3   &  $A^\prime$                          &   & 1349.3  & 1324.9   & $A^\prime$            \\
 9    & 1130.5  &  1109.3   &  $A^\prime$                          &   & 1102.4  & 1085.1   & $A^\prime$            \\
10    &  868.0  &   853.5   &  $A^\prime$                          &   &  877.4  &  860.8   & $A^\prime$            \\
11    &  661.2  &   656.0   &  $A^\prime$                          &   &  666.7  &  656.6   & $A^\prime$            \\
12    &  452.9  &   453.1   &  $A^\prime$                          &   &  495.8  &  494.3   & $A^\prime$            \\
13    &  138.7  &   136.5   &  $A^\prime$                          &   &  114.1  &  110.1   & $A^\prime$            \\
14    &  995.6  &   972.1   &  $A^{\prime\prime}$                  &   & 1010.6  &  987.8   & $A^{\prime\prime}$    \\
15    &  932.3  &   915.6   &  $A^{\prime\prime}$                  &   &  920.9  &  903.8   & $A^{\prime\prime}$    \\
16    &  710.5  &   698.4   &  $A^{\prime\prime}$                  &   &  703.0  &  690.9   & $A^{\prime\prime}$    \\
17    &  603.4  &   601.1   &  $A^{\prime\prime}$                  &   &  608.8  &  607.8   & $A^{\prime\prime}$    \\
18    &   80.9  &    78.3   &  $A^{\prime\prime}$                  &   &   98.2  &   90.0   & $A^{\prime\prime}$    \\
\vspace{-0.3cm}\\
\hline
\end{tabular}
\end{footnotesize}
\end{center}
\end{table*}

\begin{table*}[!h]
\caption{List of the measured transitions of \textit{trans} vinyl isocyanate.}
\label{transitions-trans}
\begin{center}
\begin{footnotesize}
\setlength{\tabcolsep}{4pt}
\begin{tabular}{r r r r r r r r c c r r r}
\hline\hline
\vspace{-0.3cm}\\
 $J'$ & $K_{a}'$  & $K_{c}'$ & $J''$ & $K_{a}''$ & $K_{c}''$ & $\nu_{\text{obs}}$ (MHz) \tablefootmark{a} & $\nu_{\text{obs}}-\nu_{\text{calc}}$ (MHz) \tablefootmark{b} & $u_{\text{obs}}$ (MHz) \tablefootmark{c}  & $(\nu_{\text{obs}}-\nu_{\text{calc}})_{\text{blends}}$ (MHz) \tablefootmark{d} & Weight \tablefootmark{e} & Notes\tablefootmark{f}\\
\vspace{-0.3cm}\\
\hline
\vspace{-0.3cm}\\
 18 & 0 & 18 &  17 & 1 & 17 &   32486.6530 & --0.0066  &  0.030 &           &       &(1) \\
  7 & 0 &  7 &   6 & 0 &  6 &   33483.4190 &   0.0144  &  0.030 &           &       &(1) \\
 27 & 4 & 23 &  26 & 4 & 22 &  129188.7710 & --0.0446  &  0.020 & --0.0009  & 0.50  &(2) \\
 27 & 4 & 24 &  26 & 4 & 23 &  129188.7710 &   0.0428  &  0.020 & --0.0009  & 0.50  &(2) \\
 28 & 3 & 25 &  27 & 3 & 24 &  133995.4910 &   0.0153  &  0.020 &           &       &(2) \\
 40 & 2 & 38 &  39 & 2 & 37 &  192144.7701 &   0.0190  &  0.020 &           &       &(2) \\
\vspace{-0.3cm}\\
\hline
\end{tabular}
\end{footnotesize}
\end{center}
\tablefoot{
\tablefoottext{a}{Observed frequency.}
\tablefoottext{b}{Observed minus calculated frequency.}
\tablefoottext{c}{Uncertainty of the observed frequency.}
\tablefoottext{d}{Observed minus calculated frequency for blends.}
\tablefoottext{e}{Intensity weighting factor for blended transitions.}
\tablefoottext{f}{Source of the data: (1) \cite{Kirby1978}, (2) This work.}
This table is available in its entirety in electronic form at the CDS via anonymous ftp to cdsarc.u-strasbg.fr (130.79.128.5) or via
http://cdsweb.u-strasbg.fr/cgi-bin/qcat?J/A+A/. A portion is shown here for guidance regarding its form and content.}
\end{table*}

\begin{table*}[!h]
\caption{List of the measured transitions of \textit{cis} vinyl isocyanate.}
\label{transitions-cis}
\begin{center}
\begin{footnotesize}
\setlength{\tabcolsep}{4pt}
\begin{tabular}{r r r r r r r r c c r r r}
\hline\hline
\vspace{-0.3cm}\\
 $J'$ & $K_{a}'$  & $K_{c}'$ & $J''$ & $K_{a}''$ & $K_{c}''$ & $\nu_{\text{obs}}$ (MHz) \tablefootmark{a} & $\nu_{\text{obs}}-\nu_{\text{calc}}$ (MHz) \tablefootmark{b} & $u_{\text{obs}}$ (MHz) \tablefootmark{c}  & $(\nu_{\text{obs}}-\nu_{\text{calc}})_{\text{blends}}$ (MHz) \tablefootmark{d} & Weight \tablefootmark{e} & Notes\tablefootmark{f}\\
\vspace{-0.3cm}\\
\hline
\vspace{-0.3cm}\\
  5 & 0 &  5 &   4 & 0 &  4 &   28831.8370 &   0.0116  &  0.050 &           &       &(1) \\
  5 & 3 &  2 &   4 & 3 &  1 &   29017.1090 & --0.0433  &  0.050 &           &       &(1) \\
  3 & 0 &  3 &   2 & 0 &  2 &   17360.0500 &   0.0996  &  0.100 &           &       &(2) \\
 28 & 1 & 27 &  27 & 1 & 26 &  159197.4792 & --0.0176  &  0.030 &           &       &(3) \\
 31 &13 & 18 &  30 &13 & 17 &  180427.4988 &   0.0182  &  0.030 &  0.0183   & 0.50  &(3) \\
 31 &13 & 19 &  30 &13 & 18 &  180427.4988 &   0.0182  &  0.030 &  0.0183   & 0.50  &(3) \\
\vspace{-0.3cm}\\
\hline
\end{tabular}
\end{footnotesize}
\end{center}
\tablefoot{
\tablefoottext{a}{Observed frequency.}
\tablefoottext{b}{Observed minus calculated frequency.}
\tablefoottext{c}{Uncertainty of the observed frequency.}
\tablefoottext{d}{Observed minus calculated frequency for blends.}
\tablefoottext{e}{Intensity weighting factor for blended transitions.}
\tablefoottext{f}{Source of the data: (1) \cite{Kirby1978}, (2) \cite{Bouchy1977}, (3) This work.}
This table is available in its entirety in electronic form at the CDS via anonymous ftp to cdsarc.u-strasbg.fr (130.79.128.5) or via
http://cdsweb.u-strasbg.fr/cgi-bin/qcat?J/A+A/. A portion is shown here for guidance regarding its form and content.}
\end{table*}

\begin{table*}
\caption{Rotational partition functions for \textit{cis} vinyl isocyanate with the $0_{0,0}$ energy level set to 0~cm$^{-1}$.}
\label{part-fce-cisE0}
\begin{center}
\begin{footnotesize}
\setlength{\tabcolsep}{8.0pt}
\begin{tabular}{ r r }
\hline\hline
\vspace{-0.3cm}\\
$T$ (K) &  $Q_{\text{rot}}$  \\
\vspace{-0.3cm}\\
\hline
\vspace{-0.3cm}\\
   300.000    &  67984.09   \\
   225.000    &  44103.25  \\
   150.000    &  23971.62   \\
    75.000    &   8461.98  \\
    37.500    &   2990.40  \\
    18.750    &   1057.91  \\
     9.375    &    374.79  \\
     5.000    &    146.55  \\
     2.725    &     59.38 \\
\hline
\end{tabular}
\end{footnotesize}
\end{center}
\end{table*}

\begin{table*}[!h]
\caption{JPL/CDMS catalog line list for \textit{trans} vinyl isocyanate.}
\label{predictions-trans}
\begin{center}
\begin{footnotesize}
\setlength{\tabcolsep}{6pt}
\begin{tabular}{r r r r r r r r c c r r r r}
\hline\hline
\vspace{-0.3cm}\\
$\nu_{\text{calc}}$ (MHz) \tablefootmark{a} & $u_{\text{calc}}$ (MHz) \tablefootmark{b} & Log(Int) \tablefootmark{c} & DR \tablefootmark{d} & $E$$_{\rm low}$ (cm$^{-1}$)\tablefootmark{e}& $g_{\rm upp}$ \tablefootmark{f} &
TAG \tablefootmark{g} & QNFMT \tablefootmark{h}
          & $J'$ & $K_{a}'$  & $K_{c}'$  &  $J''$ & $K_{a}''$  & $K_{c}''$  \\
\vspace{-0.3cm}\\
\hline
\vspace{-0.3cm}\\
    4784.2235  & 0.0002  & --7.9908  & 3  &    0.0000 &   3  &  69520  & 303 &  1  & 0 &  1   &      0 &  0 &  0   \\
    5121.9296  & 0.0028  & --8.0323  & 3  &   10.5300 &  21  &  69520  & 303 & 10  & 1 & 10   &     11 &  0 & 11   \\
    5439.2173  & 0.0030  & --7.9046  & 3  &   14.3358 &  27  &  69520  & 303 & 13  & 0 & 13   &     12 &  1 & 12   \\
    5933.7122  & 0.0169  & --8.0585  & 3  &  263.6818 & 111  &  69520  & 303 & 55  & 3 & 53   &     56 &  2 & 54   \\
    7077.2898  & 0.0183  & --7.9264  & 3  &  281.9165 & 117  &  69520  & 303 & 58  & 2 & 56   &     57 &  3 & 55   \\
    8426.1650  & 0.0073  & --7.5472  & 3  &   97.5187 &  69  &  69520  & 303 & 34  & 1 & 33   &     33 &  2 & 32   \\
    9477.2707  & 0.0003  & --7.2256  & 3  &    2.1659 &   5  &  69520  & 303 &  2  & 1 &  2   &      1 &  1 &  1   \\
    9568.3370  & 0.0003  & --7.0882  & 3  &    0.1596 &   5  &  69520  & 303 &  2  & 0 &  2   &      1 &  0 &  1   \\
    9659.7256  & 0.0003  & --7.2091  & 3  &    2.1689 &   5  &  69520  & 303 &  2  & 1 &  1   &      1 &  1 &  0   \\
\hline
\end{tabular}
\end{footnotesize}
\end{center}
\tablefoot{
\tablefoottext{a}{Predicted frequency.}
\tablefoottext{b}{Predicted uncertainty of the frequency.}
\tablefoottext{c}{Base 10 logarithm of the integrated intensity at 300\,K in units of nm$^2$\,MHz.}
\tablefoottext{d}{Degrees of freedom in the rotational partition function.}
\tablefoottext{e}{Lower state energy.}
\tablefoottext{f}{Upper state degeneracy.}
\tablefoottext{g}{Species tag or molecular identifier.}
\tablefoottext{h}{Format of the quantum numbers.}
The key parameters used in the generation of this table: $\mu_{a} = 2.047$~D, $\mu_{b} = 0.824$~D, $T = 300$~K, and $Q_{\text{rot}} = 62500.3370$
which takes into account the ground vibrational states of both isomers.
Predictions of \textit{a}-type transitions should be reliable up to $J=68$ for $K_a=0-5$ and $J=44$ for $K_a=6$, which basically
correspond to the data sets encompassed in this work and interpolations between them. Predictions with $K_a > 6$ are not recommended for use and $K_a < 6$ beyond 330 GHz should
be viewed with caution.
This table is available in its entirety in electronic form at the CDS via anonymous ftp to cdsarc.u-strasbg.fr (130.79.128.5) or via
http://cdsweb.u-strasbg.fr/cgi-bin/qcat?J/A+A/. A portion is shown here for guidance regarding its form and content. The table will be also available in CDMS.}
\end{table*}

\begin{table*}[!h]
\caption{JPL/CDMS catalog line list for \textit{cis} vinyl isocyanate.}
\label{predictions-cis}
\begin{center}
\begin{footnotesize}
\setlength{\tabcolsep}{6pt}
\begin{tabular}{r r r r r r r r c c r r r r}
\hline\hline
\vspace{-0.3cm}\\
$\nu_{\text{calc}}$ (MHz) \tablefootmark{a} & $u_{\text{calc}}$ (MHz) \tablefootmark{b} & Log(Int) \tablefootmark{c} & DR \tablefootmark{d} & $E$$_{\rm low}$ (cm$^{-1}$)\tablefootmark{e}& $g_{\rm upp}$ \tablefootmark{f} &
TAG \tablefootmark{g} & QNFMT \tablefootmark{h}
          & $J'$ & $K_{a}'$  & $K_{c}'$  &  $J''$ & $K_{a}''$  & $K_{c}''$  \\
\vspace{-0.3cm}\\
\hline
\vspace{-0.3cm}\\
    5796.8672  &  0.0003  & --8.4124 &  3  &  301.0000 &   3  &  69521 &  303 &  1  & 0 &  1   &      0 &  0 &  0  \\
    7479.2457  &  0.0168  & --8.3913 &  3  &  346.8931 &  41  &  69521 &  303 & 20  & 3 & 17   &     20 &  3 & 18  \\
    8809.6192  &  0.0203  & --8.2725 &  3  &  394.6910 &  59  &  69521 &  303 & 29  & 4 & 25   &     29 &  4 & 26  \\
    9226.1922  &  0.0179  & --8.3194 &  3  &  320.8431 &  27  &  69521 &  303 & 13  & 2 & 11   &     13 &  2 & 12  \\
    9275.5117  &  0.0283  & --8.2947 &  3  &  459.3935 &  77  &  69521 &  303 & 38  & 5 & 33   &     38 &  5 & 34  \\
    9661.1586  &  0.0206  & --8.2081 &  3  &  350.9637 &  43  &  69521 &  303 & 21  & 3 & 18   &     21 &  3 & 19  \\
   11111.8730  &  0.0244  & --8.1076 &  3  &  400.5261 &  61  &  69521 &  303 & 30  & 4 & 26   &     30 &  4 & 27  \\
   11166.5595  &  0.0472  & --8.2802 &  3  &  550.3038 &  97  &  69521 &  303 & 48  & 6 & 42   &     48 &  6 & 43  \\
   11176.0408  &  0.0007  & --7.6679 &  3  &  301.7616 &   5  &  69521 &  303 &  2  & 1 &  2   &      1 &  1 &  1  \\
\hline
\end{tabular}
\end{footnotesize}
\end{center}
\tablefoot{
\tablefoottext{a}{Predicted frequency.}
\tablefoottext{b}{Predicted uncertainty of the frequency.}
\tablefoottext{c}{Base 10 logarithm of the integrated intensity at 300\,K in units of nm$^2$\,MHz.}
\tablefoottext{d}{Degrees of freedom in the rotational partition function.}
\tablefoottext{e}{Lower state energy.}
\tablefoottext{f}{Upper state degeneracy.}
\tablefoottext{g}{Species tag or molecular identifier.}
\tablefoottext{h}{Format of the quantum numbers.}
The key parameters used in the generation of this table: $\mu_{a} = 2.14$~D, $\mu_{b} = 0.09$~D, $T = 300$~K, $Q_{\text{rot}} = 62500.3370$
which takes into account the ground vibrational states of both isomers, and $E$ = 301~cm$^{-1}$. This table is available in its entirety in electronic form at the CDS via anonymous ftp to cdsarc.u-strasbg.fr (130.79.128.5) or via
http://cdsweb.u-strasbg.fr/cgi-bin/qcat?J/A+A/. A portion is shown here for guidance regarding its form and content. The table will be also available in CDMS.}
\end{table*}

\end{appendix}

\end{document}